\documentclass[11pt,a4paper]{article}

\usepackage[T1]{fontenc}
\usepackage[utf8]{inputenc}
\usepackage[margin=2.5cm]{geometry}
\usepackage{graphicx}
\usepackage{amsmath,amssymb,amsthm,mathtools,bm}
\usepackage[numbers,sort&compress]{natbib}
\usepackage[hidelinks]{hyperref}
\theoremstyle{definition}
\newtheorem{definition}{Definition}

\newcommand{\ii}{\mathrm{i}}
\newcommand{\dd}{\mathrm{d}}
\newcommand{\Ai}{\mathrm{Ai}}
\newcommand{\Kloc}{\mathcal{K}}
\newcommand{\IF}{\omega_{\mathrm{IF}}}

\DeclareMathOperator{\ImPart}{Im}
\DeclareMathOperator{\RePart}{Re}
\DeclareMathOperator{\sgn}{sgn}
\DeclareMathOperator{\Reg}{Reg}
\DeclareMathOperator{\arctanh}{arctanh}

\begin{document}

\title{Chirpons: elementary local phase oscillations associated with complex zeros}

\author{
  \begin{minipage}{0.95\textwidth}
  \centering
  Enrique G Neyra$^{1,*}$, Laureano A Bulus Rossini$^1$,\\[0.3em] Pablo A Costanzo Caso$^1$, Marcelo F Ciappina$^{4,5,6}$ and Lorena Reb\'on$^{2,3}$\\[1em]
  {\small
$^1$Instituto Balseiro, Universidad Nacional de Cuyo and Comisi\'on Nacional de Energ\'ia At\'omica, and CONICET CCT Patagonia Norte, Av.\ Bustillo 9500, 8400 San Carlos de Bariloche, R\'io Negro, Argentina\\[0.4em]
$^2$Departamento de F\'isica, Facultad de Ciencias Exactas, Universidad Nacional de La Plata, C.C.\ 67, 1900 La Plata, Argentina\\[0.4em]
$^3$Instituto de F\'isica de La Plata, UNLP--CONICET, 1900 La Plata, Argentina\\[0.4em]
$^4$Department of Physics, Guangdong Technion--Israel Institute of Technology, 241 Daxue Road, Shantou, Guangdong 515063, People's Republic of China\\[0.4em]
$^5$Technion--Israel Institute of Technology, Haifa 32000, Israel\\[0.4em]
$^6$Guangdong Provincial Key Laboratory of Materials and Technologies for Energy Conversion, Guangdong Technion--Israel Institute of Technology, 241 Daxue Road, Shantou, Guangdong 515063, People's Republic of China\\[0.6em]
$^*$Author to whom any correspondence should be addressed.\\
\texttt{\href{mailto:enrique.neyra@ib.edu.ar}{enrique.neyra@ib.edu.ar}}
  }
  \end{minipage}
}
\date{}
\maketitle

\begin{abstract}
For a complex wave, the local phase variation provides the instantaneous frequency in time, the local wavenumber in space, and the local momentum in quantum mechanics. Here we study this quantity from the complex-zero structure of the analytic continuation of the wave. A single nonreal zero produces a localized Lorentzian phase variation on the real axis, whose position, width, sign, amplitude, and integrated phase are fixed by the zero itself. We call this elementary zero-associated oscillation a \emph{chirpon}. The representation is exact for finite products and finite Fourier sums, and extends to entire and band-limited functions through canonical factorization. We illustrate the construction with superoscillatory sequences and discrete Fourier sums, showing how local anomalous oscillations can be reconstructed from periodic families of complex zeros. We further show how linear systems transform chirpons through the displacement of their associated complex zeros, and illustrate this dynamics for quadratic dispersive propagation and a resonant harmonic oscillator. This establishes a direct relation between local phase structure and the amplitude and phase response of the transfer function, suggesting a zero-based form of spectroscopy in which system parameters can be inferred from the motion, width, and orientation of chirpons. More broadly, the chirpon description provides a unified framework for instantaneous frequency, local wavenumber, superoscillation, backflow, and other local phase phenomena, while offering a new route for probing linear systems through the dynamics of their complex zeros.
\end{abstract}


\section{Introduction}
\label{sec:introduction}

A complex wave can be written locally as $F(\xi)=A(\xi)\exp[\ii\theta(\xi)]$, with $A(\xi)>0$, where $\xi$ denotes a real observation variable. The quantity
\begin{equation}
  \Kloc_\xi(\xi)=\frac{\dd\theta}{\dd\xi}=\ImPart\left[\frac{\partial_\xi F(\xi)}{F(\xi)}\right]
  \label{eq:general-local-phase}
\end{equation}
measures the local variation of the phase. Although its mathematical form is universal, its physical interpretation depends on the variable represented by $\xi$. For a temporal analytic signal it is the instantaneous angular frequency, a central quantity in analytic-signal and time--frequency theory \cite{Gabor1946,Boashash1992,Picinbono1997,Flandrin2019}. For a scalar wave in space, the corresponding phase gradient defines the local wavevector, normal to the constant-phase surfaces \cite{BerryShukla2019}.

The same quantity has a direct quantum-mechanical interpretation. Writing the wavefunction locally as $\psi(\bm r,t)=R(\bm r,t)\exp[\ii S(\bm r,t)/\hbar]$, the spatial phase gradient determines the local momentum through the gradient of the action $S$. In the Madelung formulation it determines the probability-flow velocity, while in the de Broglie--Bohm formulation it gives the corresponding guidance velocity in the absence of a vector potential \cite{Madelung1927,Bohm1952}. The same local momentum also appears naturally in weak-measurement theory: when momentum is weakly measured and position is postselected, the real part of the weak momentum is precisely the momentum associated with the local phase gradient \cite{Aharonov1988,Wiseman2007,Berry2013}. Optical weak-measurement experiments have provided an operational reconstruction of such local momentum fields \cite{Kocsis2011}.

The behavior of the local phase gradient becomes especially striking when local and global spectral information disagree. In a band-limited wave, a region where the local oscillation rate exceeds the largest Fourier component is superoscillatory. In space this corresponds to an anomalously large local wavevector, while in the weak-value language it can be interpreted as an anomalous weak momentum \cite{BerryPopescu2006,Aharonov2017,BerryShukla2021,ZheludevYuan2022}. A complementary phenomenon is quantum backflow: a wavefunction composed only of positive momenta can nevertheless develop regions of negative probability current. In the local-phase picture this reversal is associated with a local phase gradient opposite to the globally allowed momentum direction. Closely related reverse-flow phenomena also occur in optical fields \cite{BrackenMelloy1994,Berry2010,ZheludevYuan2022}.

More generally, singularities are ubiquitous geometrical structures in wave fields, arising whenever some property of the chosen wave description becomes locally degenerate or undefined. Depending on the level of description, they appear as caustics in ray optics, phase singularities or wavefront dislocations in scalar waves, and polarization singularities in vector fields \cite{Berry2000,SoskinVasnetsov2001,Berry2023,angelsky2022optical}. Large and rapidly varying local phase gradients are intimately connected with zeros of the complex field. A zero on the physical domain is a phase singularity: the amplitude vanishes and the phase itself becomes undefined. In its neighborhood, however, the phase can vary extremely rapidly, producing large local wavevectors and superoscillatory regions \cite{NyeBerry1974,BerryShukla2019,BerryShukla2021,ZheludevYuan2022,grant2025localization}.

The connection between zeros, phase, and modulation has a substantial history in signal theory. Voelcker emphasized real and complex zeros as fundamental attributes of band-limited signals and developed a description of modulation in terms of zero manipulation \cite{Voelcker1966a,Voelcker1966b}. Picinbono investigated the constraints relating instantaneous amplitude and phase, including the role of Blaschke factors \cite{Picinbono1997,Picinbono1998}. Related representations in terms of complex zeros, Blaschke products, and inner functions have since been used to analyze instantaneous frequency and to characterize analytic phase signals and mono-components \cite{RaoKumaresan1997,KumaresanRao1999,KumaresanRao2000,QianChen2006,ChenLiWu2020}.

In a series of previous works, we progressively developed the physical phenomena that motivate the present formulation. Destructive interference between waves with different characteristic widths was first used to generate experimentally localized sub-diffraction structures in the spatial domain and sub-Fourier ultrashort pulses in the temporal domain \cite{neyra2021tailoring,neyra2022simple}. We subsequently introduced the concept of super-bandwidth (SB), showing that such a wavepacket can interact locally with a physical system as if it possessed an effective bandwidth larger than that inferred from its global Fourier spectrum \cite{Neyra21,neyra2022principal}. In the zero-based language developed here, these interference constructions can be viewed as controlling field zeros and thereby creating localized regions with anomalous wave structure.

We later studied the propagation of SB pulses through a dispersive medium and found that atypical local oscillations emerge in the vicinity of the field zeros, with instantaneous frequencies extending beyond the spectral range characterizing the incident pulse \cite{neyra2024superbandwidth}. Exploiting the mathematical equivalence between dispersive optical propagation and the free evolution of a quantum wavepacket, we subsequently found the corresponding anomalous behavior in the local momentum of a freely propagating SB wavepacket, including superoscillatory and suboscillatory regimes \cite{neyra2026free}. Most recently, we introduced the term \emph{chirpon} for the localized oscillation emerging from a one-dimensional phase singularity after interaction with a linear system, and verified this behavior experimentally using a resonant system \cite{neyra2025chirpons}.

The purpose of the present work is to identify the analytic structure underlying these observations and to formulate the chirpon independently of any particular wavepacket or physical system. After analytic continuation of the observation variable, an individual nonreal zero produces a pole in the logarithmic derivative of the field and, consequently, a localized variation of the phase along the real axis. We identify this elementary zero-associated contribution as a \emph{chirpon}. Depending on the physical variable under consideration, the same mathematical structure represents a contribution to the instantaneous frequency in time, to the local wavevector in space, or, after multiplication by $\hbar$, to the quantum local momentum. This viewpoint provides a common language for these apparently different local oscillatory phenomena while keeping their physical interpretations distinct.

Beyond this local definition, we develop representations of the local phase rate in terms of complex zeros and show how prescribed zero patterns can be used to synthesize oscillatory structures with finite Fourier sums. We then investigate how linear systems transform chirpons through the displacement of their associated zeros, deriving local approximations that relate this motion to the transfer function and the field surrounding each zero. These expressions explain previously reported theoretical and experimental observations, including the response of super-bandwidth wavepackets to resonant and dispersive systems. The sinc and finite-energy Airy wavepackets provide complementary examples that clarify the role of the incident zero structure and the scope of the approximations. Together, these results suggest a route to characterizing linear systems through the evolution of chirpon positions, widths, and orientations.

\section{Chirpons as local phase oscillations associated with complex zeros}
\label{sec:definition}

Let $F(z)$ be analytic in a domain containing a portion of the real axis. On the real axis, wherever $F(\xi)\neq0$, it can be written as $F(\xi)=A(\xi)e^{\ii\theta(\xi)}$, with $A(\xi)=|F(\xi)|$. Its logarithmic derivative is therefore $F'(\xi)/F(\xi)=A'(\xi)/A(\xi)+\ii\,\dd\theta/\dd\xi$, so that the local phase rate is given directly by $\dd\theta/\dd\xi=\ImPart[F'(\xi)/F(\xi)]$, in agreement with equation~\eqref{eq:general-local-phase}.

This representation makes the role of the complex zeros of $F$ particularly transparent. Consider a simple nonreal zero $z_n=\xi_n+\ii\gamma_n$, with $\gamma_n\neq0$. In a neighborhood of $z_n$ the function can be factorized as $F(z)=(z-z_n)G_n(z)$, where $G_n(z)$ is analytic and nonzero near the zero, $G_n(z_n)\neq0$. Taking the logarithmic derivative gives
\begin{equation}
  \frac{F'(z)}{F(z)}=\frac{1}{z-z_n}+\frac{G_n'(z)}{G_n(z)}.
  \label{eq:local-logder}
\end{equation}
Thus, a simple zero generates a simple pole in the logarithmic derivative, while $G_n'/G_n$ provides a regular background contribution.

Restricting equation~\eqref{eq:local-logder} to the real axis, the contribution associated specifically with the zero is $1/(\xi-z_n)=(\xi-\xi_n+\ii\gamma_n)/[(\xi-\xi_n)^2+\gamma_n^2]$. Its imaginary part is therefore
\begin{equation}
  \ImPart\left[\frac{1}{\xi-z_n}\right]=\frac{\gamma_n}{(\xi-\xi_n)^2+\gamma_n^2}.
  \label{eq:elementary-phase-rate}
\end{equation}
Equivalently, the elementary zero factor on the real axis, $\xi-z_n=(\xi-\xi_n)-\ii\gamma_n$, has amplitude $A_n(\xi)=|\xi-z_n|=[(\xi-\xi_n)^2+\gamma_n^2]^{1/2}$ and phase $\theta_n(\xi)=\arg(\xi-z_n)$, whose derivative is precisely equation~\eqref{eq:elementary-phase-rate}.

\begin{definition}[Chirpon]
A \emph{chirpon} is the elementary local phase oscillation associated with a single nonreal complex zero $z_n=\xi_n+\ii\gamma_n$ of an analytic function. Its contribution to the local phase rate is
\begin{equation}
  \chi_n(\xi)=\frac{\gamma_n}{(\xi-\xi_n)^2+\gamma_n^2}.
  \label{eq:chirpon-definition}
\end{equation}
The center $\xi_n$ determines the position of the oscillation, while $|\gamma_n|$ sets its characteristic width and the sign of $\gamma_n$ determines the sign of the phase excursion. When $\xi=t$, $\chi_n$ is an elementary instantaneous-angular-frequency contribution; when $\xi=x$, it is an elementary local-wavenumber contribution.
\end{definition}

The chirpon therefore arises directly from the pole structure of the logarithmic derivative $F'/F$: each nonreal zero of $F$ contributes a localized Lorentzian term to the derivative of its phase, superimposed on the regular phase variation associated with the nonvanishing background.

The geometry of the zero completely determines the properties of the chirpon. Its real coordinate $\xi_n$ fixes its center, while $|\gamma_n|$, the distance from the real axis, determines its width and peak magnitude. At the center $\chi_n(\xi_n)=1/\gamma_n$, so that $|\chi_{n,\max}|=1/|\gamma_n|$. The full width at half maximum is $\mathrm{FWHM}=2|\gamma_n|$, and therefore $|\chi_{n,\max}|\,\mathrm{FWHM}=2$. Zeros closer to the real axis thus produce narrower and stronger local phase variations, whereas more distant zeros produce broader and weaker ones. The sign is determined by the half-plane containing the zero: $\gamma_n>0$ produces a positive chirpon and $\gamma_n<0$ a negative one.

The same parameter also determines the amplitude of the elementary zero factor, $A_n(\xi)$, whose minimum occurs at the center, $A_{n,\min}=|\gamma_n|$. Hence the minimum amplitude and the maximum chirpon magnitude satisfy the simple relation $A_{n,\min}|\chi_{n,\max}|=1$. Thus, as a zero approaches the real axis, the associated amplitude minimum decreases while the local phase-rate excursion increases in exactly inverse proportion.

The integrated phase variation is independent of the distance from the real axis, $\int_{-\infty}^{\infty}\chi_n(\xi)\,\dd\xi=\pi\sgn(\gamma_n)$. Vertical motion of the zero therefore changes the width, amplitude minimum, and peak phase rate, but preserves the total signed phase variation. A simple nonreal zero carries one elementary phase contribution of magnitude $\pi$.

Finally, as the zero approaches the real axis, the chirpon becomes increasingly narrow and its peak diverges while its area remains fixed. In the distributional sense,
\begin{equation}
  \lim_{\gamma\rightarrow0^\pm}\frac{\gamma}{(\xi-\xi_n)^2+\gamma^2}=\pm\pi\delta(\xi-\xi_n).
  \label{eq:real-zero-limit}
\end{equation}
A zero on the observation axis is therefore the singular limit of a finite-width chirpon and corresponds to a phase jump at an amplitude zero, providing the one-dimensional counterpart of the connection between zeros and phase singularities in wave fields \cite{NyeBerry1974}.

\subsection{Zeros with multiplicity}

Let $z_n=\xi_n+\ii\gamma_n$ be a zero of multiplicity $M\in\mathbb N$, so that locally $F(z)=(z-z_n)^M G_n(z)$ with $G_n(z_n)\neq0$. Since $(z-z_n)^M=\prod_{j=1}^{M}(z-z_n)$, the multiple zero can be viewed as $M$ coincident simple zeros. It therefore represents $M$ identical chirpons,
\begin{equation}
  \chi_n^{(M)}(\xi)=M\chi_n(\xi)=M\frac{\gamma_n}{(\xi-\xi_n)^2+\gamma_n^2}.
  \label{eq:multiplicity-chirpon}
\end{equation}
Multiplicity scales the phase-rate excursion without changing the center or the width: $|\chi_{n,\max}^{(M)}|=M/|\gamma_n|$ and $\mathrm{FWHM}=2|\gamma_n|$, so that $|\chi_{n,\max}^{(M)}|\,\mathrm{FWHM}=2M$. The elementary amplitude associated with the multiple zero is $A_n^{(M)}(\xi)=|\xi-z_n|^M=[(\xi-\xi_n)^2+\gamma_n^2]^{M/2}$, with minimum $A_{n,\min}^{(M)}=|\gamma_n|^M$. Hence $|\chi_{n,\max}^{(M)}|=M\,[A_{n,\min}^{(M)}]^{-1/M}$. The total phase contribution is $\int_{-\infty}^{\infty}\chi_n^{(M)}(\xi)\,\dd\xi=M\pi\sgn(\gamma_n)$. Thus the multiplicity is not an arbitrary continuous weight: it counts the number of coincident elementary zero contributions.

\section{General form of the local phase rate}
\label{sec:general-form}

\subsection{Finite products}

The finite case follows directly from a polynomial. A polynomial with zeros $z_n$ of multiplicities $m_n$ can be written as $P(z)=C\prod_{n=1}^{N}(z-z_n)^{m_n}$, with $C\neq0$. More generally, including a zero-free exponential factor, $F(z)=C\exp(az)\prod_{n=1}^{N}(z-z_n)^{m_n}$, whose logarithmic derivative is $F'(z)/F(z)=a+\sum_{n=1}^{N}m_n/(z-z_n)$. Writing each nonreal zero as $z_n=\xi_n+\ii\gamma_n$, evaluation on the real axis gives
\begin{equation}
  \Kloc_\xi(\xi)=\ImPart(a)+\sum_{n=1}^{N}m_n\frac{\gamma_n}{(\xi-\xi_n)^2+\gamma_n^2}.
  \label{eq:finite-chirpon-sum}
\end{equation}
Thus, for a polynomial, for which $a=0$, the regular local phase rate is a finite sum of chirpons. Real zeros ($\gamma_n=0$) contribute no regular imaginary part away from the zero itself, but appear as singular phase jumps in the limit described by equation~\eqref{eq:real-zero-limit}.

\subsection{Entire functions and canonical products}

A holomorphic function is locally represented by a convergent power series. If it is holomorphic over the whole complex plane, it is an \emph{entire} function. For entire functions with infinitely many zeros, the finite polynomial factorization is replaced by a convergent canonical product.

Let $F$ be an entire function of finite order, with a zero of multiplicity $m_0$ at the origin and nonzero zeros $\{z_n\}$. Its Hadamard--Weierstrass factorization may be written as \cite{Boas1954,Levin1980} $F(z)=z^{m_0}\exp[Q(z)]\prod_n E_p(z/z_n)^{m_n}$, where $Q$ is a polynomial and $E_p(w)=(1-w)\exp(w+w^2/2+\cdots+w^p/p)$ is the Weierstrass elementary factor. Taking the logarithmic derivative gives
\begin{equation}
  \frac{F'(z)}{F(z)}=\frac{m_0}{z}+Q'(z)+\sum_n m_n\left[\frac{1}{z-z_n}+\sum_{k=1}^{p}\frac{z^{k-1}}{z_n^k}\right].
  \label{eq:general-logder}
\end{equation}
Each zero therefore contributes the same elementary pole $1/(z-z_n)$, whose imaginary part on the real axis is a chirpon. The remaining terms form the regular background required by the analytic class and by convergence of the infinite product. Hence
\begin{equation}
  \Kloc_\xi(\xi)=\Kloc_{\mathrm{bg}}(\xi)+\Reg\sum_n m_n\frac{\gamma_n}{(\xi-\xi_n)^2+\gamma_n^2},
  \label{eq:general-local-phase-rate}
\end{equation}
where $\Reg$ denotes the canonical ordering and regularization inherited from the factorization. The chirpon terms retain exactly the same form as in the finite case.

\subsection{Finite-energy band-limited functions}

Consider a finite-energy band-limited function $F(\xi)$, with square-integrable Fourier spectrum $\widetilde{F}(\omega)$ supported in $[\omega_-,\omega_+]$. Its analytic continuation to the complex plane is $F(z)=(2\pi)^{-1/2}\int_{\omega_-}^{\omega_+}\widetilde{F}(\omega)\exp(\ii\omega z)\,\dd\omega$. By the Paley--Wiener theorem, this continuation is an entire function of exponential type \cite{PaleyWiener1934}. Finite-energy Paley--Wiener functions belong to the Cartwright class, for which the zeros can be ordered so that the function admits the factorization \cite{Levin1980} $F(z)=Cz^{m_0}\exp(\ii K_c z)\prod_n(1-z/z_n)^{m_n}$, where $C\neq0$, $K_c$ is a real constant associated with the zero-free exponential factor, and the infinite product is understood with the corresponding canonical ordering of the zeros. Taking the logarithmic derivative gives $F'(z)/F(z)=m_0/z+\ii K_c+\sum_n m_n/(z-z_n)$. Writing each nonreal zero as $z_n=\xi_n+\ii\gamma_n$ and evaluating the imaginary part on the real axis, the local phase rate becomes
\begin{equation}
  \Kloc_\xi(\xi)=K_c+\sum_n m_n\frac{\gamma_n}{(\xi-\xi_n)^2+\gamma_n^2},
  \label{eq:PW-chirpons-general}
\end{equation}
where, as before, real zeros (including the one at the origin) contribute only singular phase jumps. Thus, the local phase rate of a finite-energy band-limited function is a constant background plus the sum of the chirpons associated with its nonreal zeros. Band limitation constrains the allowed zero distribution and the background term, but it does not modify the elementary form of each chirpon.

For a temporal wave, $\xi=t$, the local phase rate is the instantaneous angular frequency, and equation~\eqref{eq:PW-chirpons-general} becomes $\IF(t)=\omega_c+\sum_n m_n\gamma_n/[(t-t_n)^2+\gamma_n^2]$, with $\omega_c\equiv K_c$ and $z_n=t_n+\ii\gamma_n$. This is the instantaneous-frequency representation obtained previously from the complex-zero structure of the signal. For $\xi=x$, the same general expression gives the local wavenumber. For a temporal wave, the corresponding result for the instantaneous frequency was already recognized in the classical work of Voelcker more than half a century ago \cite{Voelcker1966a,Voelcker1966b}. In the present formulation, this result acquires a direct local interpretation: each complex zero contributes one elementary chirpon to the instantaneous frequency.

More generally, the local phase rate is not a Fourier variable selected from the spectral support. It is determined by the constant background together with the geometry of the complex zeros. Consequently, band limitation does not impose a pointwise bound on the local phase rate. A zero sufficiently close to the real axis can generate a chirpon with a very large positive or negative excursion while the Fourier spectrum remains strictly confined to its original band.

This provides a unified zero-based interpretation of anomalous local oscillatory behavior. Depending on the variable and physical context, the local phase rate may exceed the upper spectral bound, fall below the lower bound, or even reverse sign. These regimes include superoscillatory and suboscillatory behavior and, when the local phase rate reverses relative to a spectrum of fixed sign, backflow. They therefore arise from the same local mechanism: the phase contributions generated by complex zeros.

\section{Examples}
\label{sec:examples}

\subsection{The canonical superoscillatory function}

Consider the standard superoscillatory sequence
\begin{equation}
  F_N(z;a)=\left[\cos\left(\frac{z}{N}\right)+\ii a\sin\left(\frac{z}{N}\right)\right]^N,\qquad a>1.
  \label{eq:Berry-function}
\end{equation}
This function is a finite trigonometric polynomial whose Fourier components are confined to the interval $[-1,1]$ in the variable conjugate to $\xi$. Nevertheless, for fixed $\xi$ and large $N$, its behavior around the origin approaches $\exp(\ii a\xi)$, so that the local phase rate approaches $a>1$. This is the canonical example of superoscillation \cite{BerryPopescu2006,Aharonov2017}.

Introducing $\eta=\arctanh(1/a)>0$, i.e.\ $a=\coth\eta$, equation~\eqref{eq:Berry-function} can be written as $F_N(z;a)=\{\ii\sin[(z-\ii N\eta)/N]/\sinh\eta\}^N$. Its zeros are $z_\ell=N\pi\ell+\ii N\eta$, $\ell\in\mathbb{Z}$, each with multiplicity $N$. Using the product representation of the sine, the symmetric pairs of zeros can be grouped as
\begin{equation}
  F_N(z;a)=\left[\frac{\ii(z-z_0)}{N\sinh\eta}\prod_{\ell=1}^{\infty}\left(1-\frac{(z-z_0)^2}{N^2\pi^2\ell^2}\right)\right]^N,\qquad z_0=\ii N\eta,
  \label{eq:Berry-zero-product}
\end{equation}
and the logarithmic derivative is $F_N'/F_N=N[1/(z-z_0)+\sum_{\ell=1}^{\infty}(1/(z-z_\ell)+1/(z-z_{-\ell}))]$. On the real axis, each pole contributes to the local phase rate through $\ImPart[N/(\xi-z_\ell)]=N^2\eta/[(\xi-N\pi\ell)^2+(N\eta)^2]\equiv\chi_\ell(\xi)$. Since the chirpon terms decay as $\ell^{-2}$, the paired series is absolutely convergent and can equivalently be written as
\begin{equation}
  \Kloc_{\xi,N}(\xi;a)=\sum_{\ell\in\mathbb{Z}}\chi_\ell(\xi)=\sum_{\ell\in\mathbb{Z}}\frac{N^2\eta}{(\xi-N\pi\ell)^2+(N\eta)^2}.
  \label{eq:Berry-chirpon-series}
\end{equation}
The central contribution, $\chi_0(\xi)=N^2\eta/[\xi^2+(N\eta)^2]$, has peak value $1/\eta$ and full width at half maximum $2N\eta$. For large $a$, $\eta\simeq1/a$, and therefore $\chi_0(0)\simeq a$.

Using the standard periodic Lorentzian sum, equation~\eqref{eq:Berry-chirpon-series} can be evaluated exactly as
\begin{equation}
  \Kloc_{\xi,N}(\xi;a)=\frac{\sinh(2\eta)}{\cosh(2\eta)-\cos(2\xi/N)}=\frac{a}{\cos^2(\xi/N)+a^2\sin^2(\xi/N)}.
  \label{eq:Berry-chirpon-sum}
\end{equation}
In particular, $\Kloc_{\xi,N}(0;a)=a$, although all Fourier components of $F_N$ lie inside $[-1,1]$. More generally, for $a>1$, $1/a\le\Kloc_{\xi,N}(\xi;a)\le a$, with maxima $a$ at $\xi=mN\pi$ and minima $1/a$ at $\xi=N\pi/2+mN\pi$, $m\in\mathbb Z$. Thus the superoscillatory phase rate does not correspond to an additional Fourier component. It emerges from the geometry of the complex-zero lattice, equivalently from the superposition of its associated chirpons.

\begin{figure}[!htbp]
  \centering
  \includegraphics[width=0.80\linewidth]{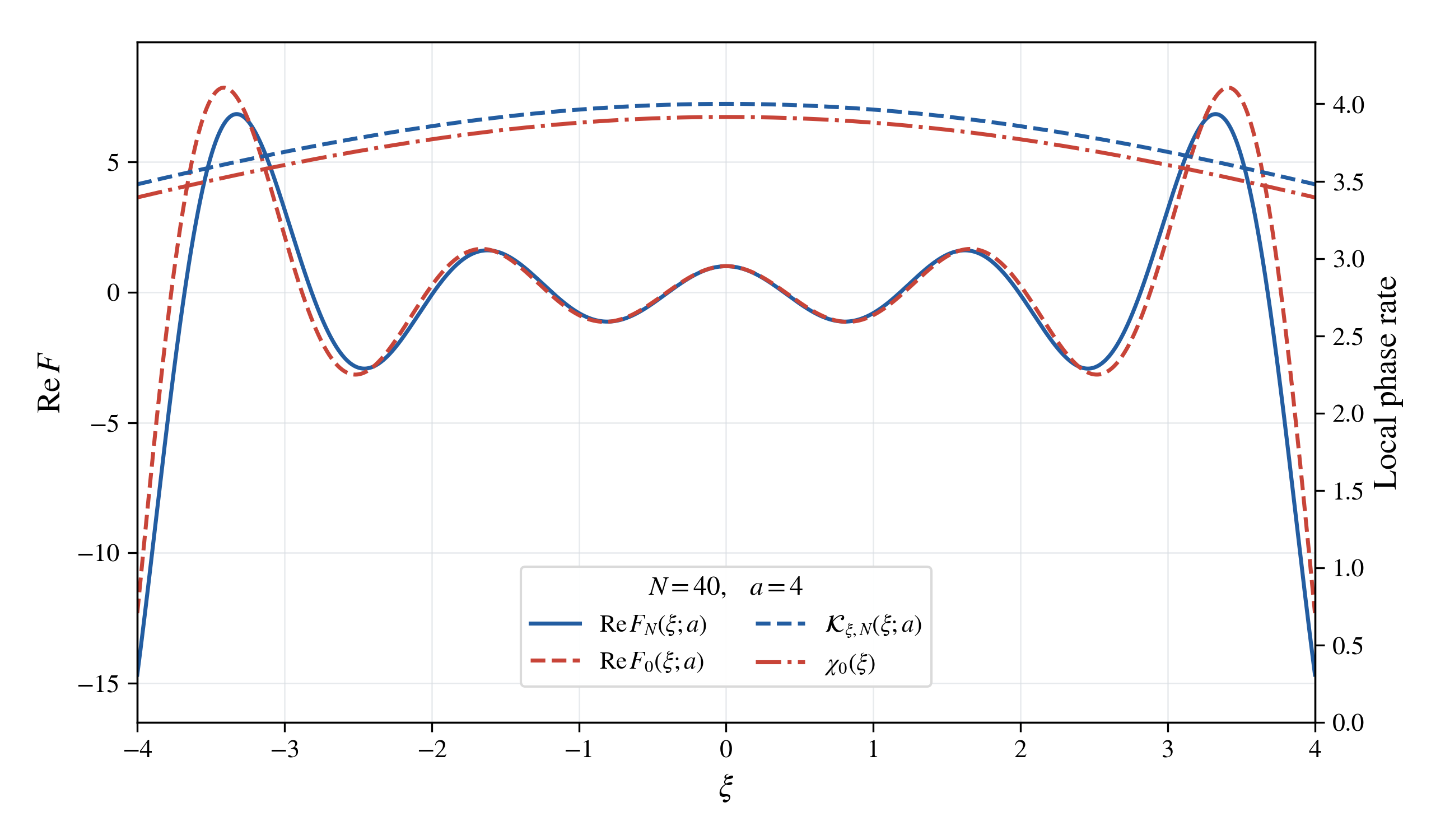}

  \vspace{0.3cm}

  \includegraphics[width=0.80\linewidth]{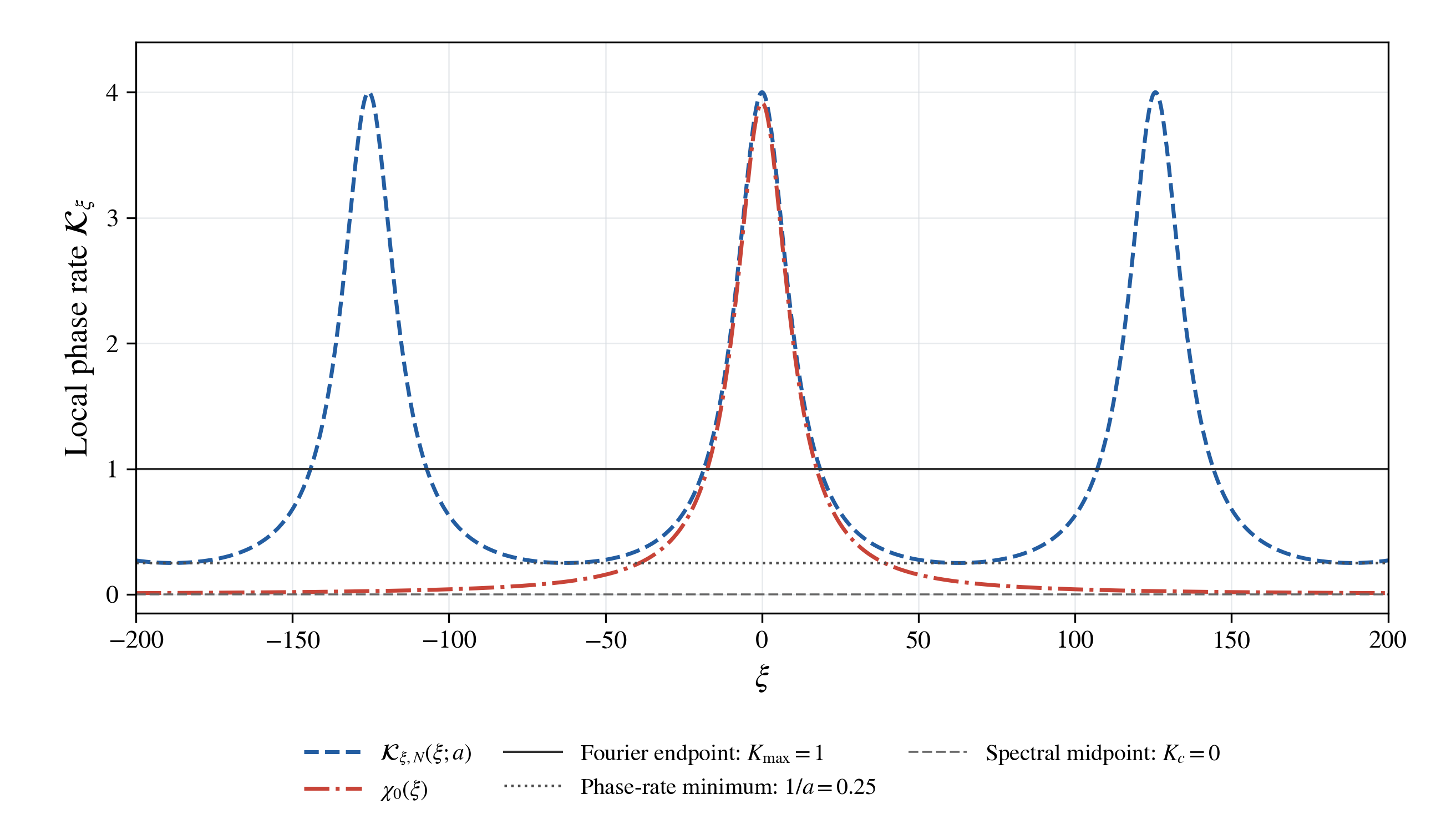}
  \caption{Canonical superoscillatory function and its representation in terms of the central chirpon for $a=4$ and $N=40$. Top: comparison between the real part of the canonical function $F_N(\xi;a)$ and the central-zero approximation $F_0(\xi;a)$, together with their corresponding local phase rates $\Kloc_{\xi,N}(\xi;a)$ and $\chi_0(\xi)$. Bottom: local phase rate of the canonical function over an extended interval compared with the central chirpon. The horizontal lines indicate the upper Fourier limit $K_{\max}=1$ and the spectral center $K_c=0$. The central chirpon reproduces the dominant local phase excursion around $\xi=0$, while the complete function contains additional periodic contributions associated with the remaining complex zeros.}
  \label{fig:berry-chirpon}
\end{figure}

Figure~\ref{fig:berry-chirpon} illustrates this interpretation. The upper panel compares the real part of the exact field $F_N(\xi;a)$ with the local approximation $F_0(\xi;a)$ associated with the central zero, together with their corresponding local phase rates. The lower panel shows the exact phase rate over a much wider interval, where the central chirpon captures the dominant local peak around $\xi=0$, while the full result reveals the periodic contributions generated by the remaining zeros of the lattice.

\subsection{Equally spaced discrete frequencies}

Consider a finite complex signal containing $d+1$ equally spaced frequencies, $S(t)=\sum_{k=0}^{d}c_k\exp[\ii(\omega_{\min}+k\Delta\omega)t]$, with $\Delta\omega>0$ and $c_0,c_d\neq0$. Introducing $w=\exp(\ii\Delta\omega t)$, the Fourier sum becomes a polynomial evaluated on the unit circle, $S(t)=\exp(\ii\omega_{\min}t)P_d(w)$, with $P_d(w)=\sum_{k=0}^{d}c_kw^k=C\prod_{j=1}^{J}(w-\zeta_j)^{m_j}$ and $\sum_{j=1}^{J}m_j=d$.

Each polynomial root $\zeta_j$ corresponds to a periodic family of complex zeros in the $t$ plane. Writing $\zeta_j=\exp(\ii\Delta\omega z_j)$, with $z_j=t_j+\ii\gamma_j$, these zeros are $z_{j,\ell}=t_j+\ell T+\ii\gamma_j$, with $T=2\pi/\Delta\omega$ and $\ell\in\mathbb Z$, where $t_j=\arg(\zeta_j)/\Delta\omega$ modulo $T$ and $\gamma_j=-\ln|\zeta_j|/\Delta\omega$. Thus, roots inside the unit circle generate positive chirpons, whereas roots outside the unit circle generate negative ones. The instantaneous phase rate can then be written directly in terms of these periodic zero families as
\begin{equation}
  \omega_{\mathrm{IF},S}(t)=\omega_{\min}+\frac{d\,\Delta\omega}{2}+\sum_{j=1}^{J}m_j\sum_{\ell\in\mathbb Z}\frac{\gamma_j}{(t-t_j-\ell T)^2+\gamma_j^2}.
  \label{eq:Fourier-chirpon-lattices}
\end{equation}
Thus, a finite set of equally spaced Fourier components is equivalently described by a finite set of polynomial roots in one fundamental complex-time interval. Each root generates a periodic lattice of complex zeros, and hence a periodic train of chirpons.

\subsection{Constructing an oscillation from \texorpdfstring{$N$}{N} complex points}

The previous construction can be reversed. Consider $N$ arbitrary complex points in one interval of length $T$, $z_j=\xi_j+\ii\gamma_j$, $j=1,\ldots,N$, with $\xi_{\mathrm{ref}}\leq\xi_j<\xi_{\mathrm{ref}}+T$ for some reference $\xi_{\mathrm{ref}}$. Repeating these points periodically generates the zero families $z_{j,\ell}=z_j+\ell T$, $\ell\in\mathbb Z$. An entire function having exactly these periodic zeros is
\begin{equation}
  F_N(z)=C\exp(\ii K_c z)\prod_{j=1}^{N}\sin\left[\frac{\pi}{T}(z-z_j)\right],
  \label{eq:N-point-construction}
\end{equation}
where $C\neq0$ and $K_c$ is an arbitrary constant background phase slope. Repeated points simply introduce the corresponding zero multiplicities. Taking the logarithmic derivative and evaluating its imaginary part on the real axis gives
\begin{equation}
  \Kloc_\xi(\xi)=K_c+\sum_{j=1}^{N}\sum_{\ell\in\mathbb Z}\frac{\gamma_j}{(\xi-\xi_j-\ell T)^2+\gamma_j^2}.
  \label{eq:N-point-chirpons}
\end{equation}
Thus, any set of $N$ chirpons can be reproduced exactly in one fundamental interval, together with identical copies repeated every $T$. The positions $\xi_j$, widths $|\gamma_j|$, signs, and multiplicities of the chirpons can be chosen independently.

The corresponding Fourier structure follows directly from the periodicity. Each sine factor contains the two exponential frequencies $\pm\pi/T$. Therefore, the product of $N$ factors contains only the $N+1$ frequencies $\omega_k=K_c-N\Delta\omega/2+k\Delta\omega$, $k=0,\ldots,N$, with $\Delta\omega=2\pi/T$. Consequently, equation~\eqref{eq:N-point-construction} can always be expanded as $F_N(\xi)=\sum_{k=0}^{N}c_k\exp(\ii\omega_k\xi)$, with coefficients $c_k$ completely determined by the chosen complex points and the constant $C$.

The equal spacing of the Fourier frequencies is therefore not an additional assumption. It follows from the periodic repetition of the zero pattern: a repetition period $T$ corresponds to a spectral spacing $\Delta\omega=2\pi/T$. Hence a finite set of $N$ arbitrarily chosen complex zeros provides two equivalent descriptions of the same oscillation: $N$ periodic chirpon families in the complex plane, or $N+1$ equally spaced Fourier components on the real axis.

\begin{figure}[!htbp]
  \centering
  \includegraphics[width=0.70\linewidth]{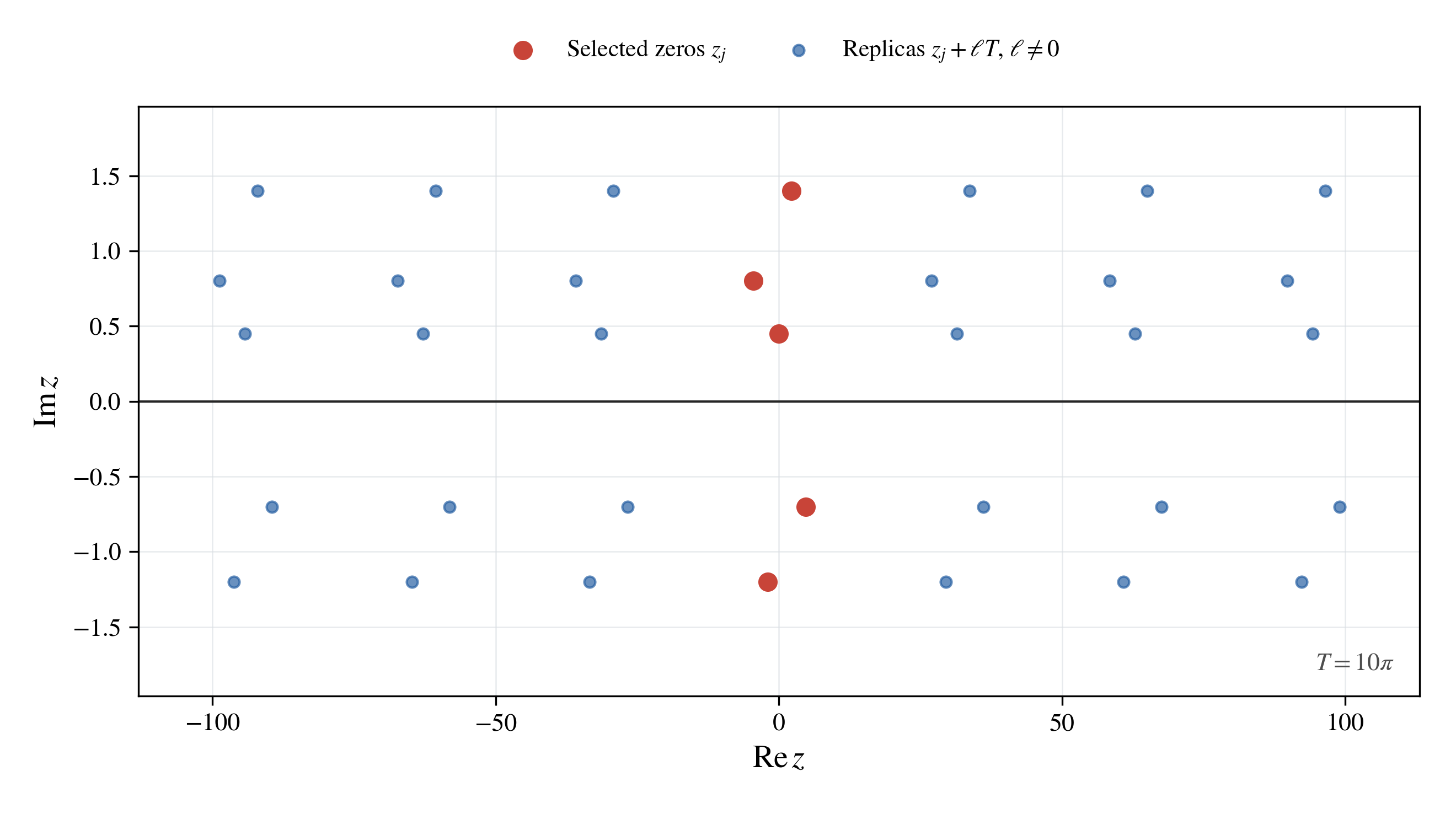}

  \vspace{0.25cm}

  \includegraphics[width=0.70\linewidth]{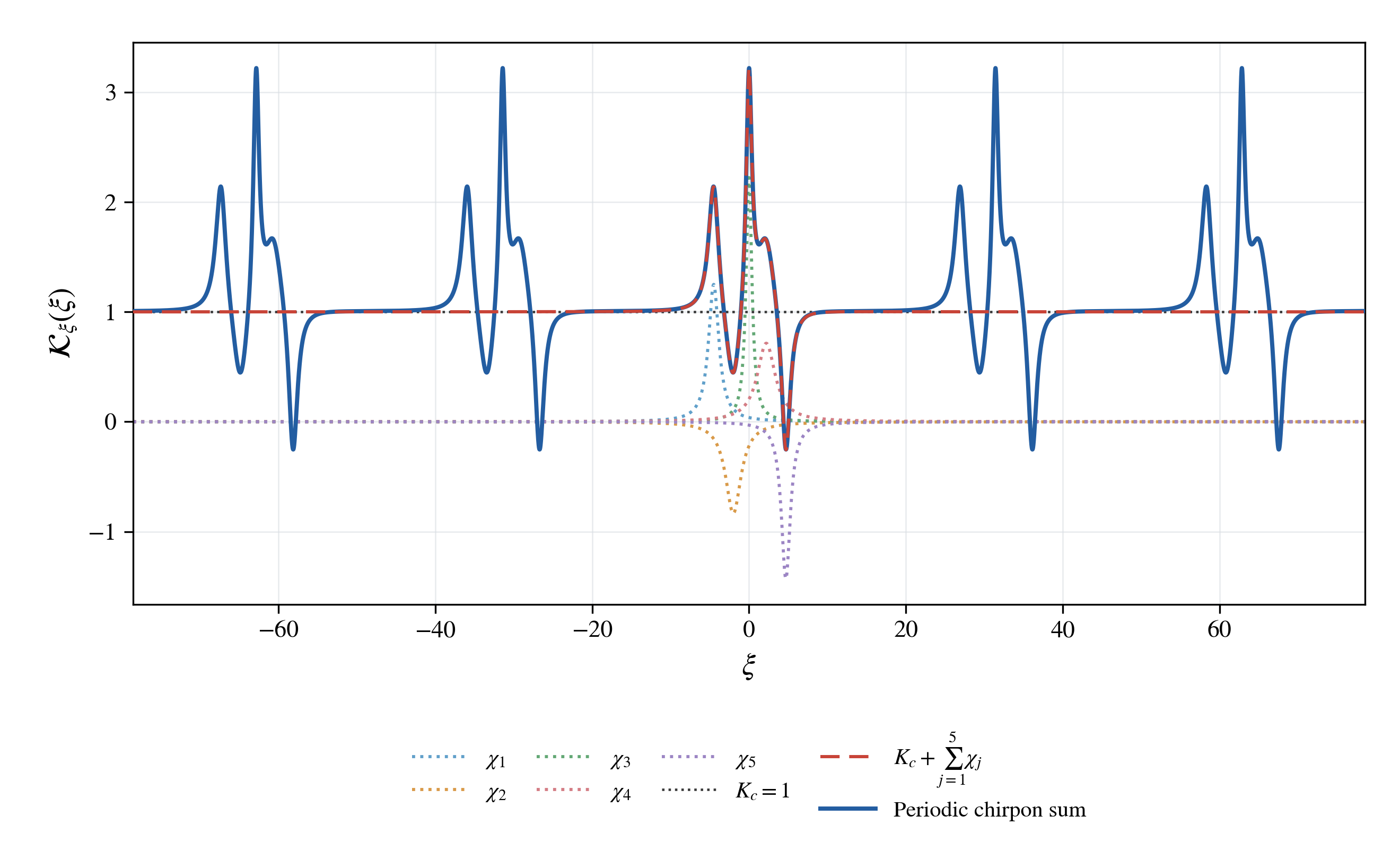}

  \vspace{0.25cm}

  \includegraphics[width=0.70\linewidth]{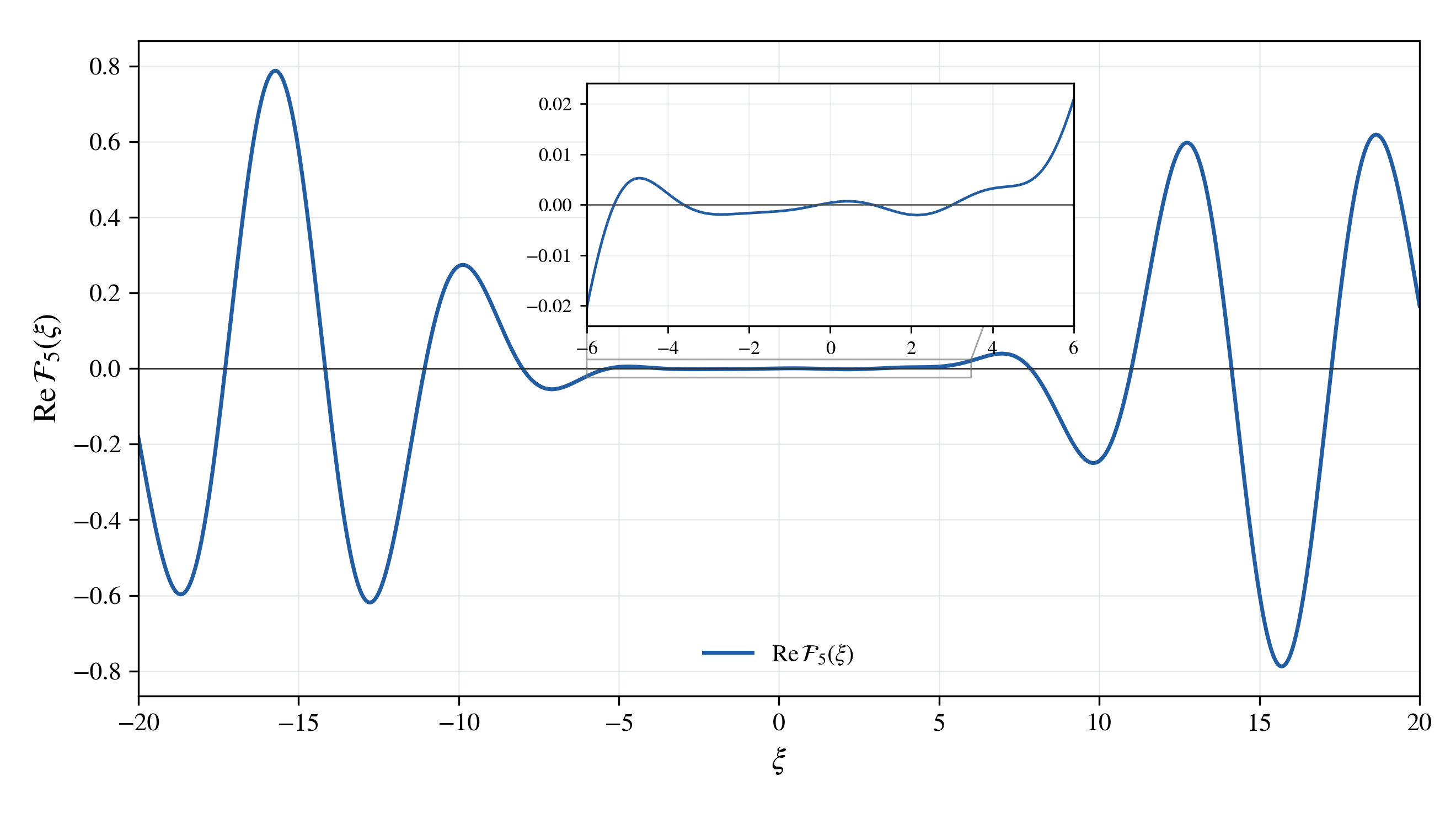}
  \caption{Construction of a finite Fourier oscillation from a prescribed set of $N=5$ complex zeros. Top: the selected zeros in one fundamental interval and their periodic replicas generated with period $T$. Middle: the individual chirpons, their finite sum, and the exact periodic chirpon lattice corresponding to the finite Fourier representation. Bottom: real part of the resulting Fourier field, with an inset showing the central region in greater detail. The construction shows the equivalence between a prescribed set of complex zeros, its periodic chirpon representation, and a finite set of $N+1$ equally spaced Fourier frequencies.}
  \label{fig:N-chirpons}
\end{figure}

This correspondence is illustrated in figure~\ref{fig:N-chirpons} for a representative set of $N=5$ prescribed complex zeros, $z_1=-4.5+0.8\ii$, $z_2=-2.0-1.2\ii$, $z_3=0.45\ii$, $z_4=2.2+1.4\ii$, and $z_5=4.7-0.7\ii$. The upper panel shows these zeros within one fundamental interval together with their periodic replicas, which define the corresponding zero lattice. The middle panel presents the individual chirpon contributions, their sum, and the exact periodic chirpon structure associated with the finite Fourier field. The lower panel shows the real part of the resulting oscillation, together with a zoom of the central region, where the local waveform is more clearly resolved. The figure therefore provides a direct visual connection between the prescribed complex-zero pattern, its periodic chirpon representation, and the equivalent finite Fourier expansion with $N+1$ equally spaced frequencies.

\section{Interaction of a chirpon with linear systems}
\label{sec:linear}

We consider an input field whose analytic continuation contains an isolated complex zero $z_0=\xi_0+\ii\gamma_0$, with $\gamma_0\neq0$. After extracting a reference carrier of angular frequency $\omega_0$, the field can be written locally as $F_{\mathrm{in}}(z)=e^{\ii\omega_0 z}(z-z_0)G(z)$, where $G(z)$ is analytic and nonzero in a neighborhood of $z_0$. Expanding the zero-free background around the zero, $G(z)=G_0[1+\mu(z-z_0)+\mathcal{O}((z-z_0)^2)]$, with $G_0=G(z_0)$ and
\begin{equation}
  \mu=\frac{G'(z_0)}{G(z_0)}.
  \label{eq:mu-definition}
\end{equation}
Along the real observation axis, the local field is therefore $F_{\mathrm{in}}(\xi)\simeq G_0e^{\ii\omega_0\xi}P(\xi)$, with $P(\xi)=(\xi-z_0)[1+\mu(\xi-z_0)]$.

Let the field propagate through a linear time-invariant system with transfer function $H(\omega)$. Direct evaluation of the convolution, given in appendix~\ref{app:linear-zero-transformation}, leads to
\begin{equation}
  F_{\mathrm{out}}(\xi)\simeq G_0H(\omega_0)e^{\ii\omega_0\xi}\left[P(\xi-\delta_H)-\mu\Lambda_H\right],
  \label{eq:linear-background-output}
\end{equation}
where
\begin{equation}
  \delta_H=\ii\frac{H'(\omega_0)}{H(\omega_0)}=\ii\,\partial_\omega\ln H\big|_{\omega_0},\qquad
  \Lambda_H=\partial_\omega^2\ln H\big|_{\omega_0}=\left[\frac{H''}{H}-\left(\frac{H'}{H}\right)^2\right]_{\omega_0}.
  \label{eq:deltaH-LambdaH}
\end{equation}
The position of the output chirpon is determined by the zero of the bracketed factor, $P(z_{\mathrm{out}}-\delta_H)-\mu\Lambda_H=0$. Introducing $y=z_{\mathrm{out}}-z_0-\delta_H$, the zero condition becomes $y+\mu y^2-\mu\Lambda_H=0$. The branch continuously connected to $y=0$ when $\Lambda_H\rightarrow0$ gives
\begin{equation}
  z_{\mathrm{out}}=z_0+\delta_H+\frac{\sqrt{1+4\mu^2\Lambda_H}-1}{2\mu}.
  \label{eq:zout-linear-background-exact}
\end{equation}
For $|\mu^2\Lambda_H|\ll1$, this expression reduces to
\begin{equation}
  z_{\mathrm{out}}\simeq z_0+\delta_H+\mu\Lambda_H.
  \label{eq:zout-linear-background}
\end{equation}

This result separates two contributions to the displacement of the complex zero. The first, $\delta_H=\ii\,\partial_\omega\ln H$, is present even for a locally constant background and represents the direct action of the linear system on the isolated chirpon. The second, $\mu\Lambda_H=[G'(z_0)/G(z_0)]\,\partial_\omega^2\ln H$, is the leading correction produced by the combined effect of the local variation of the surrounding field and the spectral curvature of the system.

In the constant-background limit, $\mu\rightarrow0$, the transformation reduces to
\begin{equation}
  z_{\mathrm{out}}=z_0+\ii\frac{H'(\omega_0)}{H(\omega_0)}.
  \label{eq:zout-constant-background}
\end{equation}
Taking the imaginary part of equation~\eqref{eq:zout-constant-background} gives $\gamma_{\mathrm{out}}=\gamma_0+\gamma_H$, where $\gamma_H=\ImPart[\ii H'(\omega_0)/H(\omega_0)]=\RePart[H'(\omega_0)/H(\omega_0)]$. This recovers, within the constant-background approximation, the empirical composition rule $\gamma_{\pm}''\simeq\gamma_{\pm}+\gamma_{\pm}'$ introduced in our previous work \cite{neyra2025chirpons}. The addition or subtraction of magnitudes observed there thus follows from the relative signs of the initial imaginary position and the system-induced displacement of the zero.

\subsection{Single-pole system}

Consider first a transfer function containing a single simple pole, $H(\omega)=C_H/(\omega-p)$, with $p=\Omega_p+\ii\kappa$. Its logarithmic derivatives are $H'/H=-1/(\omega-p)$ and $\partial_\omega^2\ln H=1/(\omega-p)^2$. Thus
\begin{equation}
  z_{\mathrm{out}}\simeq z_0-\frac{\ii}{\omega_0-p}+\frac{\mu}{(\omega_0-p)^2}.
  \label{eq:zout-single-pole-linear}
\end{equation}
The first pole-dependent term, $-\ii/(\omega_0-p)$, is the displacement obtained for a locally constant background. The second term, $\mu/(\omega_0-p)^2$, is the correction produced by the local slope of the background.

Introducing the detuning $\Delta=\Omega_p-\omega_0$ and $R=\Delta^2+\kappa^2$, the dominant displacement is $\delta_H=(\kappa+\ii\Delta)/R$, whereas $\Lambda_H=(\Delta^2-\kappa^2-2\ii\Delta\kappa)/R^2$. The first term therefore determines, to leading order, both the displacement of the chirpon along the real axis and the modification of its distance from that axis. The latter determines the characteristic width and peak local frequency excursion of the chirpon. The background correction is smaller by the factor $|\mu\Lambda_H|/|\delta_H|=|\mu|/|\omega_0-p|$. Consequently, in the regime $|\mu|\ll|\omega_0-p|$ the chirpon dynamics is dominated by the first logarithmic derivative, while the second-order term provides only a small correction.

A first-order low-pass filter, $H_{\mathrm{LP}}(\omega)=\omega_{\mathrm c}/(\omega_{\mathrm c}+\ii\omega)$, is the canonical realization of this case. Its pole is $p=\ii\omega_{\mathrm c}$, so that the leading displacement is $\delta_H=(\omega_{\mathrm c}-\ii\omega_0)/(\omega_{\mathrm c}^2+\omega_0^2)$. Thus the filter moves the complex zero both parallel and perpendicular to the real $\xi$ axis. The corresponding chirpon is therefore shifted in position and modified in width. A slowly varying background adds only the perturbative correction $\mu\Lambda_H$.

\subsection{Two-pole system: harmonic oscillator}

We now consider a two-pole linear response, $H(\omega)=C_H/[(\omega-p_1)(\omega-p_2)]$, whose logarithmic derivative is $H'/H=-1/(\omega-p_1)-1/(\omega-p_2)$. For a locally constant background, $\mu=0$, the transformation of a simple zero therefore becomes
\begin{equation}
  z_{\mathrm{out}}\simeq z_0-\ii\sum_{j=1}^{2}\frac{1}{\omega_0-p_j}.
  \label{eq:zout-two-pole}
\end{equation}
This approximation is particularly appropriate for the interaction of the symmetric SB wavepackets considered here with a damped harmonic oscillator. Indeed, both numerical simulations and experimental measurements show that the two singularities acquire nearly equal chirpon parameters, indicating that the local variation of the background produces only a small correction in this case.

The transfer function of the damped harmonic oscillator is $H_{\mathrm{HO}}(\omega)=\omega_r^2/(\omega_r^2-\omega^2+\ii\Gamma\omega)$, with $\Gamma=\omega_r/Q$. Its poles are $p_\pm=\pm\Omega_p+\ii\kappa$, where $\Omega_p=(\omega_r^2-\Gamma^2/4)^{1/2}=\omega_r(1-1/4Q^2)^{1/2}$ and $\kappa=\Gamma/2=\omega_r/2Q$. Equation~\eqref{eq:zout-two-pole} then gives $z_{\mathrm{out}}-z_0\simeq-\ii[(\omega_0-\Omega_p-\ii\kappa)^{-1}+(\omega_0+\Omega_p-\ii\kappa)^{-1}]$. For an initially real zero, $z_0=t_z$, the imaginary part of the displaced zero defines the chirpon parameter
\begin{equation}
  \gamma_{\mathrm{out}}\simeq\frac{\Omega_p-\omega_0}{(\omega_0-\Omega_p)^2+\kappa^2}-\frac{\Omega_p+\omega_0}{(\omega_0+\Omega_p)^2+\kappa^2}.
  \label{eq:gamma-HO-two-poles}
\end{equation}
The first term originates from the positive-frequency pole $p_+$, whereas the second term is the contribution of the negative-frequency pole $p_-$. The corresponding characteristic local peak frequency of the chirpon is $\Kloc_\xi^{\mathrm{peak}}\simeq\omega_0+1/\gamma_{\mathrm{out}}$. Thus, even though the input singularity is associated with the carrier $\omega_0$, its displacement after the interaction contains direct information about the spectral poles of the resonant system.

It is useful to compare equation~\eqref{eq:gamma-HO-two-poles} with the numerical results obtained previously for SB wavepackets interacting with a damped harmonic oscillator \cite{neyra2025chirpons}. In what follows, $T_0=2\pi/\omega_0$ denotes the carrier period. For $Q=10$ and $\omega_r=1.5\,\omega_0$, the previous simulations yielded a chirpon peak at approximately $\Kloc_\xi^{\mathrm{peak}}\simeq1.65\,\omega_0$. For these parameters, $\Omega_p\simeq1.498\,\omega_0$ and $\kappa=0.075\,\omega_0$, and equation~\eqref{eq:gamma-HO-two-poles} gives $\gamma_{\mathrm{out}}/T_0\simeq0.249$, which corresponds to $\Kloc_\xi^{\mathrm{peak}}\simeq1.640\,\omega_0$, in close agreement with the numerical value. For the opposite detuning, $\omega_r=0.6\,\omega_0$ with the same $Q=10$, the simulations yielded a peak frequency near $0.65\,\omega_0$. In this case $\Omega_p\simeq0.599\,\omega_0$ and $\kappa=0.030\,\omega_0$, and the two-pole expression gives $\gamma_{\mathrm{out}}/T_0\simeq-0.494$, or $\Kloc_\xi^{\mathrm{peak}}\simeq0.678\,\omega_0$, again reproducing the direction and magnitude of the numerically observed frequency shift.

The same expression can be directly compared with the experimental realization of the oscillator using an RLC circuit \cite{neyra2025chirpons}. The measured system had a resonant frequency $\nu_r=8.73$~MHz and quality factor $Q\simeq4.4$. For an input centered at $\nu_0=6$~MHz, the experimental fit gave $\gamma_{\mathrm{exp}}/T_0\simeq0.262$. Equation~\eqref{eq:gamma-HO-two-poles}, using the measured values of $\nu_r$ and $Q$, predicts $\gamma_{\mathrm{out}}/T_0\simeq0.249$, corresponding to a peak local frequency $\nu_{\mathrm{peak}}\simeq9.83$~MHz, compared with the value $\nu_{\mathrm{peak}}^{\mathrm{exp}}\simeq9.64$~MHz obtained from the experimental chirpon fit. For the second experimental case, with $\nu_0=12$~MHz, the measured chirpon parameter was $\gamma_{\mathrm{exp}}/T_0\simeq-0.706$. The two-pole approximation predicts $\gamma_{\mathrm{out}}/T_0\simeq-0.619$, corresponding to $\nu_{\mathrm{peak}}\simeq8.92$~MHz, while the experimental fit gives approximately $\nu_{\mathrm{peak}}^{\mathrm{exp}}\simeq9.29$~MHz. Thus, without introducing any fitted chirpon parameter, the displacement of the complex zero predicted from the poles of the measured resonant response reproduces the observed local frequencies on both sides of the resonance.

A simpler expression is obtained when the positive-frequency pole is well isolated from the negative-frequency pole. The relevant small parameter is
\begin{equation}
  \epsilon=\frac{|\omega_0-p_+|}{|\omega_0-p_-|}=\left[\frac{(\omega_0-\Omega_p)^2+\kappa^2}{(\omega_0+\Omega_p)^2+\kappa^2}\right]^{1/2}.
  \label{eq:single-pole-condition}
\end{equation}
When $\epsilon\ll1$, the contribution of $p_-$ varies slowly over the resonant region and the zero displacement reduces to $z_{\mathrm{out}}\simeq z_0-\ii/(\omega_0-p_+)$. Accordingly, $\gamma_{\mathrm{out}}\simeq(\Omega_p-\omega_0)/[(\omega_0-\Omega_p)^2+\kappa^2]$. If, in addition, $\kappa\ll|\Omega_p-\omega_0|\ll\Omega_p$, then $\gamma_{\mathrm{out}}\simeq1/(\Omega_p-\omega_0)$, and therefore $\Kloc_\xi^{\mathrm{peak}}\simeq\omega_0+1/\gamma_{\mathrm{out}}\simeq\Omega_p$. For a high-$Q$ oscillator, $\Omega_p=\omega_r(1-1/4Q^2)^{1/2}\simeq\omega_r[1-1/(8Q^2)]$, so that $\Kloc_\xi^{\mathrm{peak}}\simeq\Omega_p\simeq\omega_r$ for $Q\gg1$.

Hence, the simple interpretation that the chirpon peak approaches the resonance frequency emerges as the isolated-pole limit of the more general two-pole transformation. When the separation is not sufficiently large, the negative-frequency pole provides a finite correction and the complete two-pole expression, equation~\eqref{eq:gamma-HO-two-poles}, gives a more accurate description.

\subsection{Quadratic spectral phase and dispersive complex-zero dynamics}
\label{sec:quadratic-dispersion}

A qualitatively different situation arises for a system with constant spectral amplitude and a quadratic spectral phase. In this case, the constant-background approximation is not sufficient to describe the chirpons generated from initially real phase singularities.

Consider first $H(\omega)=\exp\{-\ii[\phi_0+\tau(\omega-\omega_0)+\tfrac{D}{2}(\omega-\omega_0)^2]\}$, where $\tau$ is the group delay and $D$ is the group-delay dispersion. At the carrier frequency, $\ii H'/H|_{\omega_0}=\tau$, whereas $\partial_\omega^2\ln H|_{\omega_0}=-\ii D$. If the zero is assumed to lie on a locally constant background, $\mu=0$, the local transformation gives simply $z_{\mathrm{out}}\simeq z_0+\tau$. Thus, within this approximation, quadratic dispersion only translates the zero and cannot move an initially real singularity away from the real axis. Consequently, it cannot account for the finite-width chirpons observed after dispersive propagation.

The next local order is therefore required. We write the input field in the vicinity of a simple zero $z_0$ as $F_{\mathrm{in}}(\xi)\simeq G_0e^{\ii\omega_0\xi}(\xi-z_0)[1+\mu(\xi-z_0)]$, where $\mu$, defined in equation~\eqref{eq:mu-definition}, measures the local logarithmic slope of the zero-free background. The linear-background transformation, equation~\eqref{eq:zout-linear-background}, then gives
\begin{equation}
  z_{\mathrm{out}}\simeq z_0+\tau-\ii\mu D.
  \label{eq:zout-quadratic-phase}
\end{equation}
The roles of the two spectral derivatives are therefore clearly separated: the group delay $\tau$ produces a common displacement along the real axis, whereas quadratic dispersion modifies the imaginary position of the zero through the local background slope $\mu$. For an initially real zero and real $\mu$, $\xi_{\mathrm{out}}\simeq\xi_0+\tau$ and $\gamma_{\mathrm{out}}\simeq-\mu D$.

In the following, we consider three complementary examples. The first is the super-bandwidth wavepacket discussed previously, for which the local background can be tuned through the synthesis parameters. The second is a sinc wavepacket, which contains an infinite sequence of real zeros with zero-dependent background slopes. The third is a finite-energy Airy wavepacket, for which the complete dispersive evolution of all complex zeros can be obtained in closed form.

\subsubsection{Super-bandwidth wavepackets and anomalous-like response.}
\label{sec:quadratic-SB}

We first apply equation~\eqref{eq:zout-quadratic-phase} to the super-bandwidth wavepacket. For the case $\beta=0.5$, its temporal envelope can be written as \cite{neyra2024superbandwidth,neyra2025chirpons} $A_{\mathrm{SB}}(t)=\exp(-\Delta\omega^2t^2/4)-(\alpha/2)\exp(-\Delta\omega^2t^2/16)$, so that $E_{\mathrm{SB}}(t)=A_{\mathrm{SB}}(t)e^{\ii\omega_0t}$. For $0<\alpha<2$, the envelope has two symmetric real zeros, $z_{0,\pm}=\pm t_z$, with $t_z=4\sqrt{L}/(\sqrt{3}\,\Delta\omega)$ and $L\equiv\ln(2/\alpha)$. After isolating each simple zero, the local background slope takes the compact form
\begin{equation}
  \mu_\pm=\pm\frac{\sqrt{3}\,\Delta\omega}{24\sqrt{L}}\left(3-10L\right),
  \label{eq:SB-mu-beta-half}
\end{equation}
where the upper and lower signs correspond to the zeros at $+t_z$ and $-t_z$, respectively. The detailed derivation of equation~\eqref{eq:SB-mu-beta-half} is given in appendix~\ref{app:SB-background-slope}.

Consider now the quadratic dispersive system used previously to study the propagation of SB wavepackets \cite{neyra2024superbandwidth,neyra2025chirpons},
\begin{equation}
  H_D(\omega)=\exp\left[-\ii\Omega(\omega-\omega_0)^2\right].
  \label{eq:SB-dispersive-H}
\end{equation}
This corresponds to $\tau=0$ and $D=2\Omega$, and therefore $z_{\mathrm{out},\pm}\simeq\pm t_z-2\ii\Omega\mu_\pm$. The two chirpon parameters are consequently
\begin{equation}
  \gamma_\pm\simeq-2\Omega\mu_\pm=\pm\frac{\sqrt{3}\,\Omega\Delta\omega}{12\sqrt{L}}\left(10L-3\right).
  \label{eq:SB-gamma-alpha}
\end{equation}
Hence $\gamma_-=-\gamma_+$, as expected from the symmetry of the SB wavepacket.

Equation~\eqref{eq:SB-gamma-alpha} can be directly compared with the numerical results obtained previously. For the parameters used in figure~1 of \cite{neyra2025chirpons}, $\alpha=1$, $\beta=0.5$, $\Delta\omega=0.5$, and $\Omega=2$ (in units with $T_0=1$), equation~\eqref{eq:SB-mu-beta-half} gives $\mu_\pm\simeq\mp0.1704\,T_0^{-1}$. The predicted chirpon parameters are therefore $\gamma_\pm\simeq\pm0.682\,T_0$, while the previous numerical calculation yielded $\gamma_\pm\simeq\pm0.69\,T_0$, showing very good agreement with the local complex-zero approximation. The corresponding local peak frequencies follow from $\Kloc_{\xi,\pm}^{\mathrm{peak}}\simeq\omega_0+1/\gamma_\pm$, which gives approximately $\Kloc_{\xi,+}^{\mathrm{peak}}/\omega_0\simeq1.23$ and $\Kloc_{\xi,-}^{\mathrm{peak}}/\omega_0\simeq0.77$, again reproducing the extrema observed in the previous simulations.

A further consequence of equation~\eqref{eq:SB-gamma-alpha} is that the orientation of the chirpons is not determined by the sign of the quadratic dispersion alone. Instead, it is controlled by the product $\gamma_\pm\propto-\Omega\mu_\pm$. Since $\mu_\pm$ depends on the SB synthesis parameter $\alpha$, the sign of the dispersive displacement can be reversed without changing the medium. For the right zero, the inversion occurs when $\mu_+=0$, or equivalently $3-10\ln(2/\alpha)=0$. This defines the critical value
\begin{equation}
  \alpha_c=2e^{-3/10}\simeq1.48.
  \label{eq:alpha-critical}
\end{equation}
Thus, for a fixed dispersive system with $\Omega>0$, $\alpha<\alpha_c$ implies $\mu_+<0$ and $\gamma_+>0$, whereas $\alpha>\alpha_c$ implies $\mu_+>0$ and $\gamma_+<0$. The left chirpon undergoes the opposite reversal.

At the level of the local zero dynamics, changing the sign of $\mu$ is therefore equivalent to changing the sign of the quadratic dispersion, $\Omega\rightarrow-\Omega\Longleftrightarrow\mu\rightarrow-\mu$, since both transformations reverse the sign of $\gamma_\pm$. Importantly, this does not imply that the physical dispersion of the medium changes from normal to anomalous. Rather, the same medium can produce opposite local chirpon responses because the SB synthesis modifies the geometry of the field surrounding each zero.

This provides a direct interpretation of the previously observed ``anomalous-like'' behavior of SB wavepackets \cite{neyra2024superbandwidth,neyra2026free}. Thus the apparent reversal of the local dispersive response originates from the product between the quadratic spectral phase of the medium and the local background slope associated with the phase singularity. The response is therefore determined jointly by the medium and by the local complex-zero geometry of the incident wavepacket, rather than by the sign of the material dispersion alone.

\subsubsection{Sinc wavepacket: an infinite dispersive lattice of chirpons.}
\label{sec:quadratic-sinc}

A particularly transparent extension is provided by a sinc wavepacket, whose envelope contains an infinite sequence of simple real zeros. Consider $F_{\mathrm{sinc}}(\xi)=A_{\mathrm{sinc}}(\xi)e^{\ii\omega_0\xi}$, with $A_{\mathrm{sinc}}(\xi)=\sin(B\xi)/(B\xi)$, where $B$ sets the spectral width. The zeros of the envelope are $\xi_n=n\pi/B$, $n=\pm1,\pm2,\ldots$. For a simple zero, the local background slope can be obtained directly from the derivatives of the field (see appendix~\ref{app:SB-background-slope}), $\mu_n=A_{\mathrm{sinc}}''(\xi_n)/[2A_{\mathrm{sinc}}'(\xi_n)]$, which for the sinc envelope gives $\mu_n=-1/\xi_n=-B/(n\pi)$.

Under the quadratic phase of equation~\eqref{eq:SB-dispersive-H}, equation~\eqref{eq:zout-quadratic-phase} therefore predicts
\begin{equation}
  z_n(\Omega)\simeq\xi_n+\frac{2\ii\Omega}{\xi_n}=\frac{n\pi}{B}+\ii\frac{2\Omega B}{n\pi},
  \label{eq:sinc-zero-shift}
\end{equation}
that is, $\gamma_n\simeq2\Omega/\xi_n=2\Omega B/(n\pi)$. The zeros at positive and negative $\xi$ are displaced into opposite half-planes, $\gamma_{-n}=-\gamma_n$, while the magnitude of the displacement decreases as $1/|n|$.

The corresponding local phase gradient is therefore, to leading order, $\Kloc_\xi(\xi)\simeq\omega_0+\sum_{n\neq0}\gamma_n/[(\xi-\xi_n)^2+\gamma_n^2]$. Equivalently, defining $g_n=2\Omega B/(n\pi)$ for $n>0$, the symmetric pairs can be written as
\begin{equation}
  \Kloc_\xi(\xi)-\omega_0\simeq\sum_{n=1}^{\infty}\left[\frac{g_n}{(\xi-n\pi/B)^2+g_n^2}-\frac{g_n}{(\xi+n\pi/B)^2+g_n^2}\right].
  \label{eq:sinc-Kloc-pairs}
\end{equation}

The same result can be compared directly with the exact dispersive field. Since the Fourier spectrum of the sinc envelope is rectangular, propagation through equation~\eqref{eq:SB-dispersive-H} gives, apart from an irrelevant normalization factor, $A_{\mathrm{sinc},\Omega}(\xi)\propto\int_{-B}^{B}\exp(\ii\nu\xi-\ii\Omega\nu^2)\,\dd\nu$. Its exact local phase gradient can therefore be written as
\begin{equation}
  \Kloc_\xi^{\mathrm{exact}}(\xi)=\omega_0+\RePart\left[\frac{\int_{-B}^{B}\nu\,e^{\ii\nu\xi-\ii\Omega\nu^2}\,\dd\nu}{\int_{-B}^{B}e^{\ii\nu\xi-\ii\Omega\nu^2}\,\dd\nu}\right].
  \label{eq:sinc-Kloc-exact}
\end{equation}

In the weak-dispersion regime, the collective contribution of the symmetric zero pairs produces a smooth linear phase gradient around the center of the wavepacket. Expanding equation~\eqref{eq:sinc-Kloc-pairs} around $\xi=0$ and using $\sum_{n\geq1}n^{-4}=\pi^4/90$ gives $\Kloc_\xi(\xi)-\omega_0\simeq(4\Omega B^4/45)\,\xi$, which coincides with the first-order expansion of equation~\eqref{eq:sinc-Kloc-exact} in $\Omega$ and $\xi$. Thus, the smooth central chirp can be interpreted as the collective far-field contribution of the entire lattice of displaced complex zeros.

The accuracy of equation~\eqref{eq:sinc-zero-shift} is controlled locally by $\epsilon_n=2|\Omega|B^2/(n\pi)^2$. The local approximation requires $\epsilon_n\ll1$. Consequently, the first zeros are the first to depart from the perturbative prediction as the dispersion increases, whereas the more distant zeros remain accurately described over a progressively larger dispersive range. The sinc wavepacket therefore provides a useful example in which a single quadratic spectral phase generates an infinite set of chirpons with different widths, $|\gamma_n|\propto1/|n|$, while preserving the opposite orientation of the two symmetric branches.

\begin{figure}[t]
  \centering
  \includegraphics[width=\linewidth]{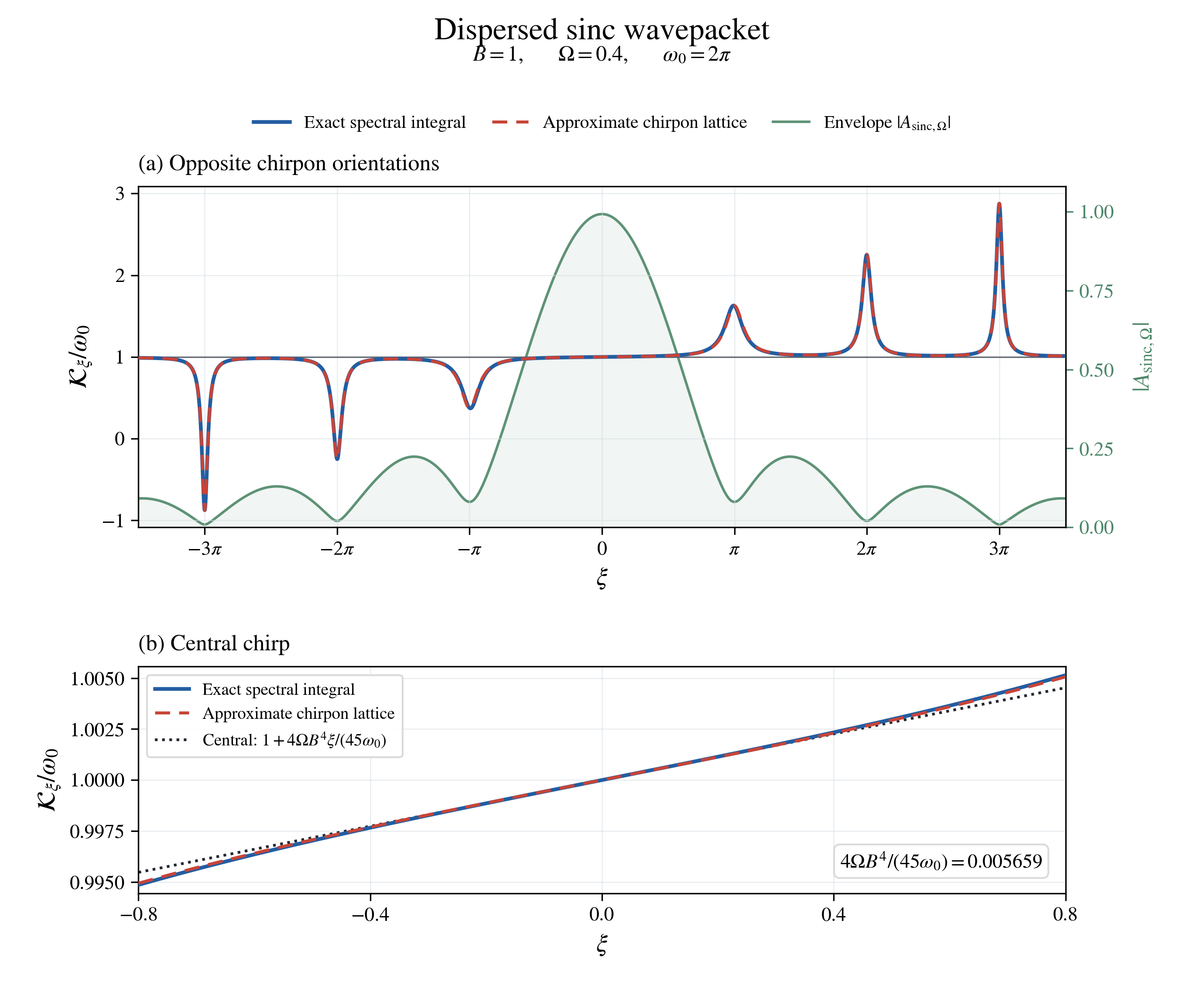}
  \caption{Dispersed sinc wavepacket for $B=1$, $\Omega=0.4$, and $\omega_0=2\pi$. (a) Normalized local phase rate $\Kloc_{\xi}/\omega_0$ obtained from the exact spectral integral (solid blue) and from the approximate chirpon lattice (dashed red). The propagated envelope modulus $|A_{\mathrm{sinc},\Omega}|$ is shown on the right axis. (b) Magnified central region, including the weak-dispersion linear approximation (dotted line).}
  \label{fig:sinc-dispersion}
\end{figure}

Figure~\ref{fig:sinc-dispersion} compares the exact local phase rate obtained from the propagated spectral integral with the approximation constructed from the displaced complex-zero lattice. The two descriptions closely agree throughout the displayed range, showing that the dispersive phase structure can be interpreted directly in terms of the chirpons generated by the sinc zeros. Since the zeros occur in symmetric pairs with opposite imaginary displacements, the corresponding chirpons have opposite orientations on the two sides of the origin. The propagated field amplitude is shown simultaneously to emphasize that the largest excursions of the local phase rate occur in the vicinity of the low-amplitude regions associated with the displaced zeros.

Near the origin, the contributions of the symmetric zero pairs combine into a smooth approximately linear phase-rate variation. In the weak-dispersion regime this yields $\Kloc_\xi/\omega_0\simeq1+4\Omega B^4\xi/(45\omega_0)$, which reproduces the central chirp of the exact solution, as shown in figure~\ref{fig:sinc-dispersion}(b). Thus, the apparently smooth central chirp and the sharp off-axis phase-rate excursions arise from the same underlying complex-zero lattice.

\subsubsection{Finite-energy Airy wavepacket: an exactly solvable zero dynamics.}
\label{sec:quadratic-airy}

A complementary example is provided by the finite-energy Airy wavepacket. In contrast to the sinc case, its complete evolution under a quadratic spectral phase can be obtained analytically. We consider the initial envelope $A_{\mathrm{Ai}}(\xi)=\Ai(\xi)e^{a\xi}$, with $a>0$, and the corresponding carrier field $F_{\mathrm{Ai}}(\xi)=A_{\mathrm{Ai}}(\xi)e^{\ii\omega_0\xi}$. For the quadratic spectral phase of equation~\eqref{eq:SB-dispersive-H}, the propagated envelope is obtained exactly \cite{zhang2015effect} as
\begin{equation}
  A_{\mathrm{Ai},\Omega}(\xi)=\Ai(q)\,e^{a\xi-2a\Omega^2}\exp\left[\ii\left(\Omega\xi+a^2\Omega-\tfrac{2}{3}\Omega^3\right)\right],\qquad q=\xi-\Omega^2+2\ii a\Omega.
  \label{eq:airy-propagated}
\end{equation}
The corresponding local phase gradient is therefore $\Kloc_\xi(\xi;\Omega)=\omega_0+\Omega+\ImPart[\Ai'(q)/\Ai(q)]$.

Let $\alpha_n<0$ denote the real simple zeros of the Airy function, $\Ai(\alpha_n)=0$. The zeros of the propagated field satisfy $q=\alpha_n$, and hence follow the exact trajectories
\begin{equation}
  z_n(\Omega)=\alpha_n+\Omega^2-2\ii a\Omega,
  \label{eq:airy-zero-trajectories}
\end{equation}
that is, $\xi_n(\Omega)=\alpha_n+\Omega^2$ and $\gamma_n(\Omega)=-2a\Omega$. A remarkable feature of this result is that all Airy zeros acquire exactly the same imaginary displacement. The quadratic spectral phase therefore generates an infinite family of chirpons with the same width and orientation, while their relative separation along the real axis remains unchanged.

Using the Hadamard product $\Ai(q)=\Ai(0)e^{bq}\prod_{n=1}^{\infty}(1-q/\alpha_n)e^{q/\alpha_n}$, where $b=\Ai'(0)/\Ai(0)$ is real, the logarithmic derivative becomes $\Ai'(q)/\Ai(q)=b+\sum_{n=1}^{\infty}[1/(q-\alpha_n)+1/\alpha_n]$. Since $b$ and $\alpha_n$ are real, the regularizing terms do not contribute to the local phase gradient. Therefore,
\begin{equation}
  \Kloc_\xi(\xi;\Omega)=\omega_0+\Omega-\sum_{n=1}^{\infty}\frac{2a\Omega}{[\xi-(\alpha_n+\Omega^2)]^2+(2a\Omega)^2}
  =\omega_0+\Omega+\sum_{n=1}^{\infty}\frac{\gamma_n}{(\xi-\xi_n)^2+\gamma_n^2}.
  \label{eq:airy-Kloc-chirpons}
\end{equation}

The Airy wavepacket also provides a direct test of the local transformation derived above. In the vicinity of any initial zero $\alpha_n$, $\mu_n=A_{\mathrm{Ai}}''(\alpha_n)/[2A_{\mathrm{Ai}}'(\alpha_n)]$. Since $\Ai''(\xi)=\xi\Ai(\xi)$, one has $\Ai''(\alpha_n)=0$, and therefore $\mu_n=a$, independently of $n$. Equation~\eqref{eq:zout-quadratic-phase} consequently predicts $z_n(\Omega)\simeq\alpha_n-2\ii a\Omega$, which reproduces exactly the term linear in $\Omega$ of equation~\eqref{eq:airy-zero-trajectories}. The exact solution further reveals the common second-order displacement along the real axis, $\Delta\xi_n=\Omega^2$.

\begin{figure}[t]
  \centering
  \includegraphics[width=\linewidth]{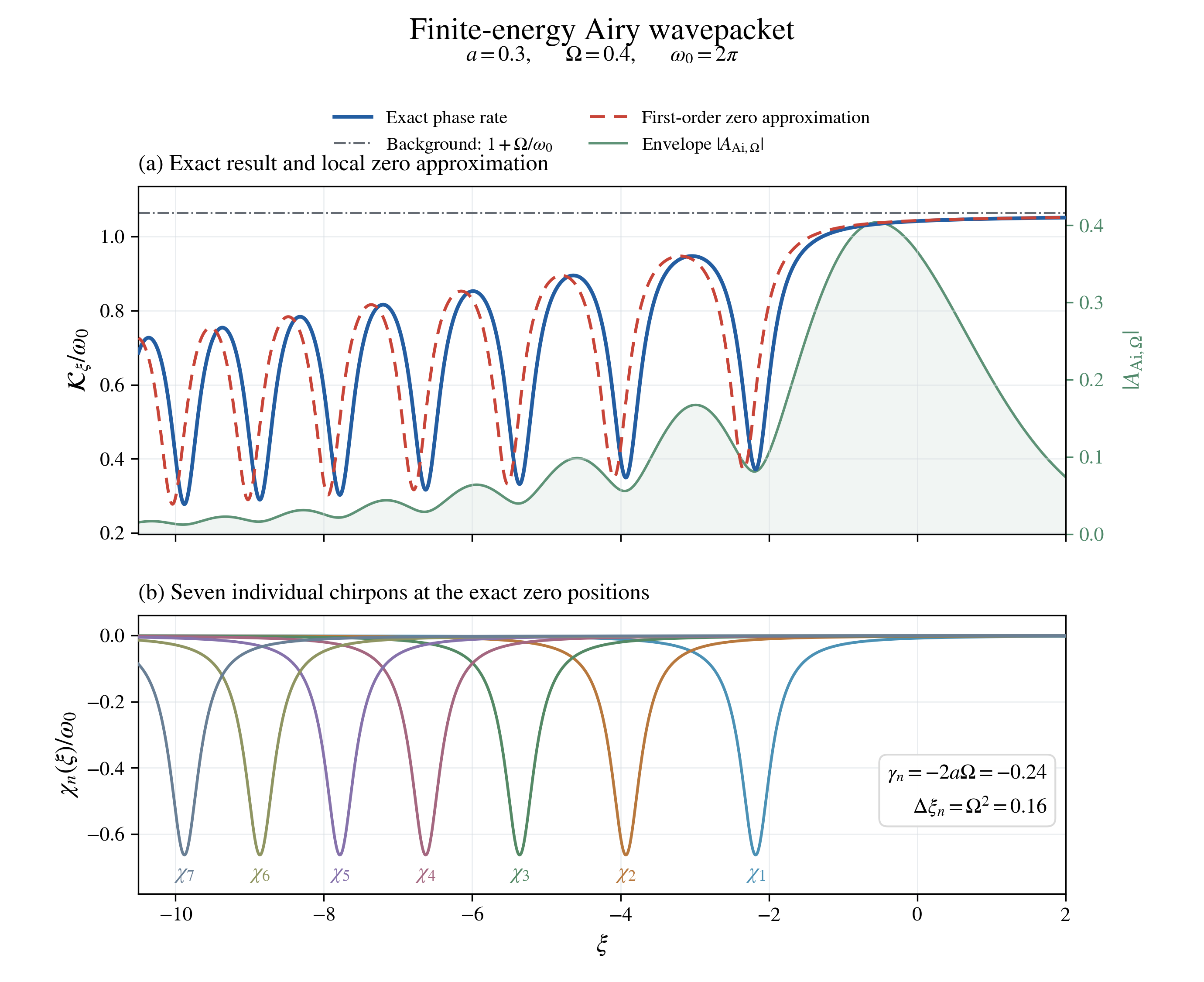}
  \caption{Finite-energy Airy wavepacket for $a=0.3$, $\Omega=0.4$, and $\omega_0=2\pi$. (a) Normalized local phase rate $\Kloc_{\xi}/\omega_0$ for the exact propagated field (solid blue) and the first-order complex-zero approximation (dashed red), together with the background contribution $1+\Omega/\omega_0$ (dash-dotted line). The propagated envelope modulus $|A_{\mathrm{Ai},\Omega}|$ is shown on the right axis. (b) Individual chirpon contributions $\chi_n(\xi)/\omega_0$ associated with the first seven complex zeros.}
  \label{fig:airy-dispersion}
\end{figure}

Figure~\ref{fig:airy-dispersion} compares the exact local phase rate of the propagated finite-energy Airy wavepacket with the approximation obtained by retaining the first-order displacement of its complex zeros. The agreement shows that the main phase structure can again be interpreted in terms of the zero dynamics induced by quadratic dispersion. In this case, all Airy zeros acquire the same imaginary displacement $\gamma_n=-2a\Omega$, while their real positions are translated by the common amount $\Omega^2$. Consequently, the associated chirpons have the same orientation and are centered at $\xi_n=\alpha_n+\Omega^2$, where $\alpha_n$ denote the zeros of the Airy function.

The individual contributions of the first seven zeros are displayed in figure~\ref{fig:airy-dispersion}(b). Their Lorentzian profiles form a nonuniform chirpon lattice whose spacing follows the Airy-zero distribution. In contrast with the sinc case, the chirpons therefore do not alternate in orientation: for the parameters considered here, $\gamma_n<0$ for every zero. The exact local phase rate additionally contains the uniform background contribution $\Omega$, originating from the linear phase term of the propagated Airy field. The close correspondence between the exact result and the zero-based approximation demonstrates that the characteristic phase-rate structure of the dispersing Airy wavepacket is directly encoded in the motion of its complex zeros.

\section{Discussion and outlook}

In this work, we have identified the contribution of a simple nonreal zero to the local phase rate as an elementary oscillation, termed a chirpon. Its position, width, peak magnitude, orientation, and integrated phase are fixed by the geometry of the zero. Its mathematical form is a Lorentzian with a parameter $\gamma$ that can be positive or negative: its magnitude sets the width, while its sign determines the orientation of the phase excursion.

Building on established zero-based descriptions of phase and modulation, this formulation connects individual zeros with the local oscillatory structure of a wavepacket. The decomposition is exact for finite products and extends to finite-energy band-limited functions through the Paley--Wiener theorem and canonical factorization. The local phase rate is thus expressed as a finite or infinite sum of chirpons, together with the appropriate background contribution. This provides a common geometric interpretation of the local phase gradients underlying superoscillation, backflow, and anomalous weak momentum.

The representative examples illustrate both the interpretation and the synthesis of these structures. In particular, prescribing a periodically repeated set of complex zeros allows the positions, widths, and orientations of the associated chirpons to be selected and realized through a finite sum of equally spaced Fourier components. This construction provides a direct route to designing local anomalous oscillations within a specified spectral framework.

We further derived simple local approximations for the evolution of complex zeros under linear systems, relating their displacement to logarithmic derivatives of the transfer function and to the field surrounding each zero. These expressions account for theoretical and experimental results reported previously, including the chirpons generated by a resonant system. A particularly revealing result concerns the reversal of the local response of a super-bandwidth wavepacket as its synthesis parameter is varied. The reversal follows from a change in the local background slope, showing how the same dispersive medium can generate opposite chirpon orientations for different incident wavepackets.

The sinc and finite-energy Airy examples further clarify this interplay between the incident field and quadratic dispersion. For the sinc wavepacket, the displaced zeros generate oppositely oriented chirpons on either side of the origin, while their collective contributions produce the smooth central chirp. For the Airy wavepacket, the exact zero trajectories reveal a common imaginary displacement and a uniform translation along the real axis, providing an explicit test of the local evolution law.

The distinction between the exact contribution of each zero and the approximate prediction of its motion sets the scope of this description. Within that scope, chirpons offer a complementary perspective on anomalous wave phenomena and suggest a route to characterizing linear systems through changes in their positions, widths, and orientations. Establishing the accuracy and robustness of parameter recovery from these observables is a natural next step toward spectroscopy based on complex-zero dynamics.


\appendix
\renewcommand{\theequation}{\Alph{section}.\arabic{equation}}

\section{Transformation of a complex zero by a linear system}
\label{app:linear-zero-transformation}
\setcounter{equation}{0}

We derive here the transformation of a complex zero produced by a linear time-invariant system directly from the convolution with its impulse response. Consider the local form of the input field, $F_{\mathrm{in}}(\xi)\simeq G_0e^{\ii\omega_0\xi}P(\xi)$, with $P(\xi)=(\xi-z_0)[1+\mu(\xi-z_0)]$, which follows from the local expansion of the zero-free background introduced in section~\ref{sec:linear}. Let $h(\eta)$ denote the impulse response of the system, with transfer function $H(\omega)=\int_{-\infty}^{\infty}h(\eta)e^{-\ii\omega\eta}\,\dd\eta$. The output field is $F_{\mathrm{out}}(\xi)=\int_{-\infty}^{\infty}h(\eta)F_{\mathrm{in}}(\xi-\eta)\,\dd\eta$, and substituting the local form of the input field,
\begin{equation}
  F_{\mathrm{out}}(\xi)=G_0e^{\ii\omega_0\xi}\int_{-\infty}^{\infty}h(\eta)e^{-\ii\omega_0\eta}P(\xi-\eta)\,\dd\eta.
  \label{eq:appendix-convolution-P}
\end{equation}

To evaluate the convolution, we define $X=\xi-z_0$, so that $P(\xi-\eta)=(X-\eta)[1+\mu(X-\eta)]=X+\mu X^2-\eta(1+2\mu X)+\mu\eta^2$. The required moments of the impulse response follow directly from the definition of $H(\omega)$. At the reference frequency $\omega_0$,
\begin{equation}
  \int_{-\infty}^{\infty}\eta^k h(\eta)e^{-\ii\omega_0\eta}\,\dd\eta=\ii^k H^{(k)}(\omega_0),\qquad k=0,1,2,
  \label{eq:appendix-moments}
\end{equation}
that is, $H(\omega_0)$, $\ii H'(\omega_0)$, and $-H''(\omega_0)$. Using equation~\eqref{eq:appendix-moments} in equation~\eqref{eq:appendix-convolution-P}, we obtain
\begin{equation}
  F_{\mathrm{out}}(\xi)=G_0e^{\ii\omega_0\xi}\left[H\left(X+\mu X^2\right)-\ii H'\left(1+2\mu X\right)-\mu H''\right]_{\omega=\omega_0}.
  \label{eq:appendix-output-intermediate}
\end{equation}
Factoring out $H(\omega_0)$ and defining $\delta_H=\ii H'(\omega_0)/H(\omega_0)$, the bracket becomes $X+\mu X^2-\delta_H(1+2\mu X)-\mu H''/H$. Since $(X-\delta_H)[1+\mu(X-\delta_H)]=X+\mu X^2-\delta_H(1+2\mu X)+\mu\delta_H^2$, this can be reorganized as $(X-\delta_H)[1+\mu(X-\delta_H)]-\mu(H''/H+\delta_H^2)$. Because $\delta_H^2=-(H'/H)^2$, we have $H''/H+\delta_H^2=H''/H-(H'/H)^2=\partial_\omega^2\ln H$, which at $\omega=\omega_0$ is the quantity $\Lambda_H$ of equation~\eqref{eq:deltaH-LambdaH}. Using $P(\xi-\delta_H)=(\xi-z_0-\delta_H)[1+\mu(\xi-z_0-\delta_H)]$, the output field takes the compact form
\begin{equation}
  F_{\mathrm{out}}(\xi)=G_0H(\omega_0)e^{\ii\omega_0\xi}\left[P(\xi-\delta_H)-\mu\Lambda_H\right].
  \label{eq:appendix-final-output}
\end{equation}

The output complex zero is determined by $P(z_{\mathrm{out}}-\delta_H)-\mu\Lambda_H=0$. Introducing $y=z_{\mathrm{out}}-z_0-\delta_H$, the explicit form of $P$ gives the quadratic equation $\mu y^2+y-\mu\Lambda_H=0$, whose solutions are $y_\pm=[-1\pm(1+4\mu^2\Lambda_H)^{1/2}]/(2\mu)$. The branch continuously connected to the original zero when the effect associated with $\Lambda_H$ vanishes is $y_+$, and hence
\begin{equation}
  z_{\mathrm{out}}=z_0+\delta_H+\frac{\sqrt{1+4\mu^2\Lambda_H}-1}{2\mu}.
  \label{eq:appendix-zout-exact}
\end{equation}
For $|\mu^2\Lambda_H|\ll1$, we may use $\sqrt{1+4\mu^2\Lambda_H}=1+2\mu^2\Lambda_H+\mathcal{O}(\mu^4\Lambda_H^2)$, which gives $z_{\mathrm{out}}\simeq z_0+\delta_H+\mu\Lambda_H$. Finally, in the constant-background limit, $\mu\rightarrow0$, the output field reduces to $F_{\mathrm{out}}(\xi)\propto\xi-z_0-\delta_H$, and therefore $z_{\mathrm{out}}=z_0+\delta_H=z_0+\ii H'(\omega_0)/H(\omega_0)$.

Thus, the first logarithmic derivative of the transfer function determines the leading complex displacement of an isolated zero, while the spectral curvature $\Lambda_H$ contributes only when the zero-free background has a nonzero local variation.

\section{Local background slope of a super-bandwidth wavepacket}
\label{app:SB-background-slope}
\setcounter{equation}{0}

Here we derive the local background slope $\mu$ associated with the two simple zeros of the super-bandwidth wavepacket considered in the main text. For $\beta=0.5$, its temporal envelope is
\begin{equation}
  A_{\mathrm{SB}}(t)=\exp\left(-\frac{\Delta\omega^2t^2}{4}\right)-\frac{\alpha}{2}\exp\left(-\frac{\Delta\omega^2t^2}{16}\right),
  \label{eq:app-SB-envelope}
\end{equation}
such that $E_{\mathrm{SB}}(t)=A_{\mathrm{SB}}(t)e^{\ii\omega_0t}$. For $0<\alpha<2$, the envelope has two symmetric real zeros located at $t=\pm t_z$. The condition $A_{\mathrm{SB}}(t_z)=0$ gives $\exp(-3\Delta\omega^2t_z^2/16)=\alpha/2$, and therefore
\begin{equation}
  t_z=\frac{4}{\sqrt{3}\,\Delta\omega}\sqrt{\ln\left(\frac{2}{\alpha}\right)}.
  \label{eq:app-SB-tz}
\end{equation}

To determine the local zero-free background, consider first the zero at $t=t_z$ and write $A_{\mathrm{SB}}(t)=(t-t_z)G_+(t)$, where $G_+(t_z)\neq0$. Expanding $G_+$ around the zero, $G_+(t)\simeq G_+(t_z)[1+\mu_+(t-t_z)]$, defines $\mu_+=G_+'(t_z)/G_+(t_z)$. For a simple zero, this quantity can be written directly in terms of the envelope: differentiating $A_{\mathrm{SB}}(t)=(t-t_z)G_+(t)$ once and twice gives $A_{\mathrm{SB}}'(t_z)=G_+(t_z)$ and $A_{\mathrm{SB}}''(t_z)=2G_+'(t_z)$, so that
\begin{equation}
  \mu_+=\frac{A_{\mathrm{SB}}''(t_z)}{2A_{\mathrm{SB}}'(t_z)}.
  \label{eq:app-mu-derivatives}
\end{equation}
The same relation, applied at any simple real zero, is used for the sinc and Airy envelopes in the main text.

For compactness, define $\lambda_1=\Delta\omega^2/4$ and $\lambda_2=\Delta\omega^2/16$, so that $A_{\mathrm{SB}}(t)=e^{-\lambda_1t^2}-(\alpha/2)e^{-\lambda_2t^2}$. Its derivatives are $A_{\mathrm{SB}}'(t)=-2\lambda_1t\,e^{-\lambda_1t^2}+\alpha\lambda_2t\,e^{-\lambda_2t^2}$ and $A_{\mathrm{SB}}''(t)=(-2\lambda_1+4\lambda_1^2t^2)e^{-\lambda_1t^2}+\alpha\lambda_2(1-2\lambda_2t^2)e^{-\lambda_2t^2}$. At $t=t_z$ the zero condition implies $e^{-\lambda_1t_z^2}=(\alpha/2)e^{-\lambda_2t_z^2}$, which yields $A_{\mathrm{SB}}'(t_z)=2t_z(\lambda_2-\lambda_1)e^{-\lambda_1t_z^2}$ and $A_{\mathrm{SB}}''(t_z)=2(\lambda_2-\lambda_1)e^{-\lambda_1t_z^2}[1-2(\lambda_1+\lambda_2)t_z^2]$. Substitution into equation~\eqref{eq:app-mu-derivatives}, with $\lambda_1+\lambda_2=5\Delta\omega^2/16$, then gives
\begin{equation}
  \mu_+=\frac{1}{2t_z}-(\lambda_1+\lambda_2)t_z=\frac{1}{2t_z}-\frac{5}{16}\Delta\omega^2t_z.
  \label{eq:app-mu-tz}
\end{equation}
Introducing $L\equiv\ln(2/\alpha)$ and substituting equation~\eqref{eq:app-SB-tz} into equation~\eqref{eq:app-mu-tz}, the result can be written exclusively in terms of the SB synthesis parameters as $\mu_+=\sqrt{3}\,\Delta\omega(3-10L)/(24\sqrt{L})$. Since $A_{\mathrm{SB}}(t)$ is an even function, $A_{\mathrm{SB}}'$ is odd and $A_{\mathrm{SB}}''$ is even, so the local background slopes at the two symmetric zeros have opposite signs, $\mu_-=-\mu_+$. Therefore,
\begin{equation}
  \mu_\pm=\pm\frac{\sqrt{3}\,\Delta\omega}{24\sqrt{L}}\left(3-10L\right),\qquad L=\ln\left(\frac{2}{\alpha}\right),
  \label{eq:app-SB-mu-final}
\end{equation}
where the upper sign corresponds to the zero at $+t_z$.

For the quadratic spectral phase $H_D(\omega)=\exp[-\ii\Omega(\omega-\omega_0)^2]$, the logarithmic derivatives evaluated at the carrier frequency are $\ii H_D'/H_D|_{\omega_0}=0$ and $\partial_\omega^2\ln H_D|_{\omega_0}=-2\ii\Omega$. The linear-background transformation derived in appendix~\ref{app:linear-zero-transformation} consequently gives $z_{\mathrm{out},\pm}\simeq\pm t_z-2\ii\Omega\mu_\pm$, and hence
\begin{equation}
  \gamma_\pm\simeq-2\Omega\mu_\pm=\pm\frac{\sqrt{3}\,\Omega\Delta\omega}{12\sqrt{L}}\left(10L-3\right).
  \label{eq:app-SB-gamma-final}
\end{equation}
Equation~\eqref{eq:app-SB-gamma-final} shows explicitly that the sign and magnitude of the dispersively generated chirpons depend jointly on the quadratic spectral phase and on the local geometry of the SB wavepacket. In particular, the sign reversal occurs at $10L-3=0$, which gives $\alpha_c=2e^{-3/10}\simeq1.48$.

\bibliographystyle{unsrt}
\bibliography{main}

\end{document}